\documentclass{aastex62}
\usepackage{amsmath,amstext,mathrsfs}
\accepted{March 17, 2019}
\usepackage{hyperref}
\usepackage{ulem}
\usepackage{natbib}
\include{bibdef}
\usepackage{lineno}

\def\et3{\eta_3}
\def\th1{\theta_{-1}}
\def\r07{r_{0,7}}
\def\x05{x_{0.5}}

\usepackage{xspace}
\newcommand{\grb}{GRB~170817A\xspace}

\newcommand{\Fermi}{\textit{Fermi\xspace}}
\newcommand{\Swift}{\textit{Swift\xspace}}
\newcommand{\INTEGRAL}{\textit{INTEGRAL\xspace}}

\newcommand{\tn}{$T_{\rm \,90}$}
\newcommand{\tf}{$T_{\rm \,50}$}
\newcommand{\tns}{$T_{\rm \,90,start}$}
\newcommand{\tfs}{$T_{\rm \,50,start}$}
\newcommand{\rb}[1]{\raisebox{1.5ex}[-1.5ex]{#1}}
\begin{document}
\title{\Fermi-GBM GRBs with characteristics similar to GRB 170817A}
\shorttitle{\Fermi-GBM GRBs similar to GRB 170817A}
\shortauthors{von Kienlin et al.}
\correspondingauthor{A.~von Kienlin}
\email{azk@mpe.mpg.de}
\author[0000-0002-0221-5916]{A.~von Kienlin}
\affiliation{Max-Planck-Institut f\"{u}r extraterrestrische Physik, Giessenbachstrasse 1, D-85748 Garching, Germany}
\author[0000-0002-2149-9846]{P.~Veres}
\affiliation{Center for Space Plasma and Aeronomic Research, University of Alabama in Huntsville, 320 Sparkman Drive, Huntsville, AL 35899, USA}
\author[0000-0002-7150-9061]{O.~J.~Roberts}
\affiliation{Science and Technology Institute, Universities Space Research Association, 320 Sparkman Drive, Huntsville, AL 35805, USA}
\author{R.~Hamburg}
\affiliation{Center for Space Plasma and Aeronomic Research, University of Alabama in Huntsville, 320 Sparkman Drive, Huntsville, AL 35899, USA}
\affiliation{Space Science Department, University of Alabama in Huntsville, 320 Sparkman Drive, Huntsville, AL 35899, USA}
\author[0000-0001-9935-8106]{E.~Bissaldi}
\affiliation{Politecnico di Bari, Via E. Orabona, 4, I-70125 Bari, Italy}
\affiliation{Istituto Nazionale di Fisica Nucleare, Sezione di Bari, I-70125 Bari, Italy}
\author{M.~S.~Briggs}
\affiliation{Center for Space Plasma and Aeronomic Research, University of Alabama in Huntsville, 320 Sparkman Drive, Huntsville, AL 35899, USA}
\affiliation{Space Science Department, University of Alabama in Huntsville, 320 Sparkman Drive, Huntsville, AL 35899, USA}
\author{E.~Burns}
\altaffiliation{NASA Postdoctoral Fellow}
\affiliation{NASA Goddard Space Flight Center, Greenbelt, MD 20771, USA}
\author[0000-0002-0587-7042]{A.~Goldstein}
\affiliation{Science and Technology Institute, Universities Space Research Association, 320 Sparkman Drive, Huntsville, AL 35805, USA}
\author{D.~Kocevski}
\affiliation{Astrophysics Office, ST12, NASA/Marshall Space Flight Center, Huntsville, AL 35812, USA}
\author[0000-0003-1626-7335]{R.~D.~Preece}
\affiliation{Center for Space Plasma and Aeronomic Research, University of Alabama in Huntsville, 320 Sparkman Drive, Huntsville, AL 35899, USA}
\affiliation{Space Science Department, University of Alabama in Huntsville, 320 Sparkman Drive, Huntsville, AL 35899, USA}
\author[0000-0002-8585-0084]{C.~A.~Wilson-Hodge}
\affiliation{Astrophysics Office, ST12, NASA/Marshall Space Flight Center, Huntsville, AL 35812, USA}
\author{C.~M.~Hui}
\affiliation{Astrophysics Office, ST12, NASA/Marshall Space Flight Center, Huntsville, AL 35812, USA}
\author{B.~Mailyan}
\affiliation{Center for Space Plasma and Aeronomic Research, University of Alabama in Huntsville, 320 Sparkman Drive, Huntsville, AL 35899, USA}
\author{C.~Malacaria}
\altaffiliation{NASA Postdoctoral Fellow}
\affiliation{NASA Marshall Space Flight Center, NSSTC, 320 Sparkman Drive, Huntsville, AL 35805, USA}
\affiliation{Universities Space Research Association, NSSTC, 320 Sparkman Drive, Huntsville, AL 35805, USA}

\begin{abstract}
We present a search for gamma-ray bursts in the \Fermi-GBM 10 yr catalog that show similar characteristics to \grb, the first electromagnetic counterpart to a GRB identified as a binary neutron star (BNS) merger 
via gravitational wave observations. Our search is focused on a nonthermal pulse, followed by a thermal component, as observed for GRB 170817A. We employ search methods based on the measured catalog parameters and Bayesian Block analysis. Our multipronged approach, which includes examination of the localization and spectral properties of the thermal component, yields a total of 13 candidates, including \grb and the previously reported similar burst, GRB 150101B. The similarity of the candidates is likely caused by the same processes that shaped the gamma-ray signal of \grb, thus providing evidence of a nearby sample of short GRBs resulting from BNS merger events.
Some of the newly identified counterparts were observed by other space telescopes and ground observatories, but none of them have a measured redshift. We present an analysis of this subsample, and we discuss two models.
From uncovering 13 candidates during a time period of ten years we predict that \Fermi-GBM will trigger on-board on about one burst similar to \grb per year.

\end{abstract}

\vspace*{1.5cm}
\section{Introduction} \label{sec:intro}
The advent of the multimessenger astronomy era was marked by the discovery of a gravitational wave (GW) signal from a neutron star merger by LIGO/VIRGO~\citep{2017ApJ...848L..12A}, followed by a short gamma-ray burst (GRB) detected by the \Fermi-gamma-ray burst monitor (\Fermi-GBM; \citealt{Goldstein+17170817}).  The \Fermi-GBM detection enabled an unprecedented, global follow-up campaign \citep{2017ApJ...848L..12A,2017ApJ...848L..13A} and unambiguously associated the electromagnetic counterpart to the GW event. While \grb showed ordinary gamma-ray properties, it had atypical rest-frame properties for a short GRB (sGRB) with a known redshift. 
As the closest sGRB with a known redshift of $z = 0.010$ (host galaxy distance;  \citealt{2017ApJ...848L..31H}), it was found to be underluminous by 3--4 orders of magnitude when compared to similar bursts. Detailed analyses of the intrinsic spectral and temporal properties of \grb revealed two components: a short ($\sim$500~ms) pulse with a nonthermal spectrum, followed by a $\sim$2~s long, soft component that is consistent with a blackbody (BB) spectrum \citep{Goldstein+17170817}. The soft tail carries around one-third of the total flux of \grb. 

Based on the importance of GRB 170817A for multimessenger astronomy, 
a search was conducted through the 10 yr \Fermi-GBM burst catalog (A. von Kienlin et al.~2019, in prep.)\footnote{GBM burst catalog entries at the \Fermi~Science Support Center (FSSC): \href{https://heasarc.gsfc.nasa.gov/W3Browse/fermi/fermigbrst.html}{https://heasarc.gsfc.nasa.gov/W3Browse/fermi/fermigbrst.html}}, namely from 2008 July 12 to 2018 July 11,  to identify  sGRBs that show similar characteristics to \grb.  \citet{2018ApJ...863L..34B} discovered GRB\,150101B, a sGRB that exhibited a short hard peak followed by a soft tail with a thermal spectrum. However, GRB\,150101B was neither subluminous nor subenergetic and occurred over a much shorter timescale than that of \grb. Various theoretical models to account for the differences in the temporal and spectral properties of these bursts are explained in detail in~\citet{2018ApJ...863L..34B}. 
Similar work exploring the fraction of already detected burst resembling \grb is described in \cite{2018Galax...6..130M} and \cite{2019MNRAS.483..840B}.

This  search will show 
the prospect of \Fermi-GBM to find nearby sGRB counterparts to GW events similar to \grb
during the upcoming LIGO/Virgo observing runs and will provide an insight to the possible range of variations of the observable characteristics. This does not inform the ability of \Fermi-GBM to detect EM counterparts for other emission scenarios of BNS or even a black hole--neutron star (BH-NS) merger event.

Further analysis of the \Fermi-GBM burst catalog using various selection criteria and methods uncovered 13 sGRBs, which include the known bursts, \grb and GRB 150101B, both of which have a hard peak followed by a soft tail. In the following, we first describe the methodology adopted to select the candidate sGRBs (Section~\ref{sec:Sec_Met}), then move on to present the major properties of the final candidate sample in comparison with the characteristics of \grb  (Section~\ref{sec:Can_Samp}). Finally, we discuss the results in the context of the theoretical models in question (Section~\ref{sec:Disc}).   

\section{Selection Methodology}
\label{sec:Sec_Met}
In order to uncover sGRBs with similar characteristics to \grb, we invoked search methods based on its most prominent observed characteristics, such as its duration and hardness ratio. Some of these quantities were derived directly from the ten year GBM burst catalog and others using a dedicated Bayesian Block analytical technique \citep{BB_scargle_2013}. 
Independent of how we finally identify the candidates, we note that we have to consider that currently only one sGRB with a GW signal has been identified and thus it is not clear if this event is typical for all such sGRBs. This raises the question of whether all BNS events exhibit a hard peak followed by a soft tail. Thus, we are not claiming completeness in the selection process presented below. We instead show that it is possible to identify similar candidates and for those bursts, present their individual properties and highlight additional common characteristics of this sample (Section \ref{sec:Can_Samp}). 

\subsection{The \Fermi-GBM instrument and data products }
\Fermi-GBM is one of two instruments on the \textit{Fermi Gamma-ray Space Telescope}, launched in 2008. GBM is made up of two types of scintillation detectors: 12 NaI(Tl) detectors, sensitive from 8 keV to $\sim$1 MeV, and two BGO detectors, sensitive from 200 keV to 40 MeV. The NaI(Tl) detectors are arranged in four groups of three on the corners of the spacecraft so that they view the whole unocculted sky. The BGO detectors are located on opposite sides of the spacecraft to enable an all-sky view. GBM produces triggered and continuous data types. A trigger is defined as a significant increase above background, detected on-board by the GBM flight software in one of several preset energy ranges and timescales \citep[for a summary of the current  settings of the GBM trigger criteria see][]{GRB_Catalog_GBM_3rd_Bhat}. Triggered data types, available since launch, include accelerated CTIME data (binned to 64 ms, 8 energy channels) and accelerated CSPEC data (binned to 1.024 s, 128 energy channels) for 10 minutes and Time Tagged Event data (individual events at 2 $\mu$s resolution, 128 energy channels) for 5 minutes after a trigger. The continuous data types used in this analysis are CTIME (256 ms, 8 energy channels, available since launch) and Continuous Time Tagged Event (CTTE) data (2 $\mu$s, 128 energy channels, available since a flight software update in 2012 November). For a complete description of the GBM instrument and data types, see \cite{Meegan09}.

\clearpage
\subsection{Sample selected using GRB catalog parameters}
\label{sec:CatParSelMeth}
Here we describe the first search we performed, which only uses parameters from the GBM GRB catalog. The 10 yr catalog (A. von Kienlin et al.~2019, in prep.) comprises 2357 GRBs, 1959 long GRBs (lGRBs, $\sim 83$\%), and 395 sGRBs ($\sim 17$\%), and 3 GRBs with no measured duration.  In order to identify those GRBs with similar temporal characteristics as \grb, we first put a constraint on the burst duration \citep[\tn, calculated in the 50--300 keV energy range, ][]{kouveliotou93} requiring that  0.5~s $<$ \tn $< $ 3.5~s. These limits were created by adding a time window of $\pm 1.5$~s around the known duration of \grb (\tn $ = 2.0 \pm 0.5$~s). This condition is fulfilled for 312 GRBs, i.e. $\sim 13$\% of GRBs of the 10 yr catalog. This is a slightly lower fraction than the one obtained using the traditional division at 2~s between the short and long GRB populations. 
Moreover, we imposed additional temporal restrictions by
employing the \tf\, duration measure as a selection tool for the brightest part of the observed GRB emission in the 50--300 keV energy range, which is mostly showing up as a main pulse. Imposing a constraint of $($\tfs$-$\tns$)/$\tn$< 0.2$ on the relative start times of \tf~ and \tn~ primarily selects GRBs with the main pulse being located at the beginning of the GRB emission, like for \grb with a relative start time difference of 0.06. Together with a second constraint on the duration ratio of  $0.1 <$ \tf$/$\tn $< 0.7$ we targeted to find GRBs where the ratio of the main pulse duration compared to the overall length is comparable to the case of \grb with \tf$/$\tn $= 0.63$ or even smaller, providing search results with a more extended tail.
In addition, we limited the 64 ms peak flux to be smaller than 10~ph~cm$^{-2}$~s$^{-1}$, thus allowing the identification of GRBs with peak fluxes up to values about twice as high as  \grb ($3.7 \pm 0.9$ ph cm$^{-2}$ s$^{-1}$). This search resulted in the identification of 107 good candidates. 

This first result highlights a relevant limitation of the search method applied. One could further restrict the parameter range to better agree with the measured values of \grb, but this would concurrently limit any variation in the observed GRB characteristics. Furthermore, the aforementioned method has a fundamental drawback; by selecting quantities which are calculated in the 50--300 keV energy band, like \tn~ and \tf, we might be missing those events where the soft component is contributing below the 50 keV threshold.

This led us to test an alternative search method, employing Bayesian Block analysis, which is described in the next Section. 

\subsection{Manual Selection Supported by Bayesian Blocks Analysis}
\label{sec:BayesBlockAnal}
We found that the most appropriate selection method is a two-step procedure, in which first GRBs with a duration of less than 5 s are prefiltered using Bayesian block analysis, followed by a dedicated manual selection. The Bayesian block analysis on its own produced a sample too large to be useful as a standalone method.

We relaxed the constraint on \tn~ used in the previous Section \ref{sec:CatParSelMeth} to \tn $< $ 5~s
driven by the wish not to exclude potential candidates with a distinct soft tail and to include short duration events like GRB~150101B. Furthermore, we dropped any constraint on the peak flux or the fluence so as to not exclude brighter events. Applying this less restricting selection criterion, we obtain a sample of 558 GRBs detected during the time period of the 10 yr trigger catalog. Aiming for the automatic detection of the main and soft emission periods of candidate GRBs, we performed Bayesian Block analysis on this sample of short GRBs over three energy channels using the triggered detectors.

The Bayesian Block analysis technique delineates a light curve into piece-wise constant Poisson rate episodes \citep{Scargle98BB}. The algorithm (applied to an interval) decides between a constant rate model and a model with two constant rates, using a change point to describe an interval. It is an iterative process that when applied, continues until a stopping condition is met. The $P$ parameter (prior) for the stopping condition indicates the chance probability of needing an additional change point. 
The analysis of \cite{BB_scargle_2013} suggests that $P \sim 10^{-2}$ is a generally a good value to use.

\begin{figure}[h!]
\centering
\begin{tabular}{c}
\includegraphics[width=0.5\textwidth]{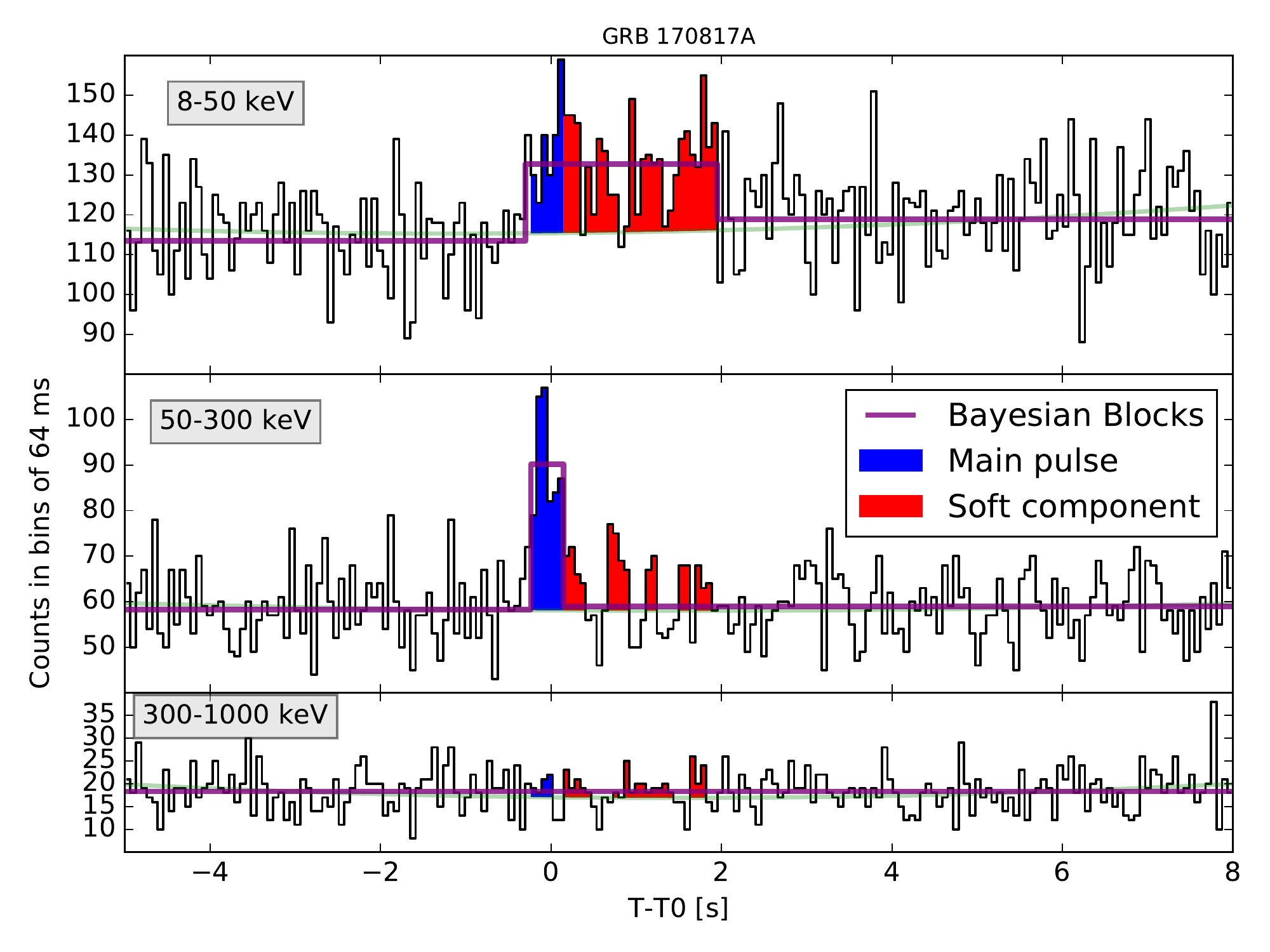} 
\includegraphics[width=0.5\textwidth]{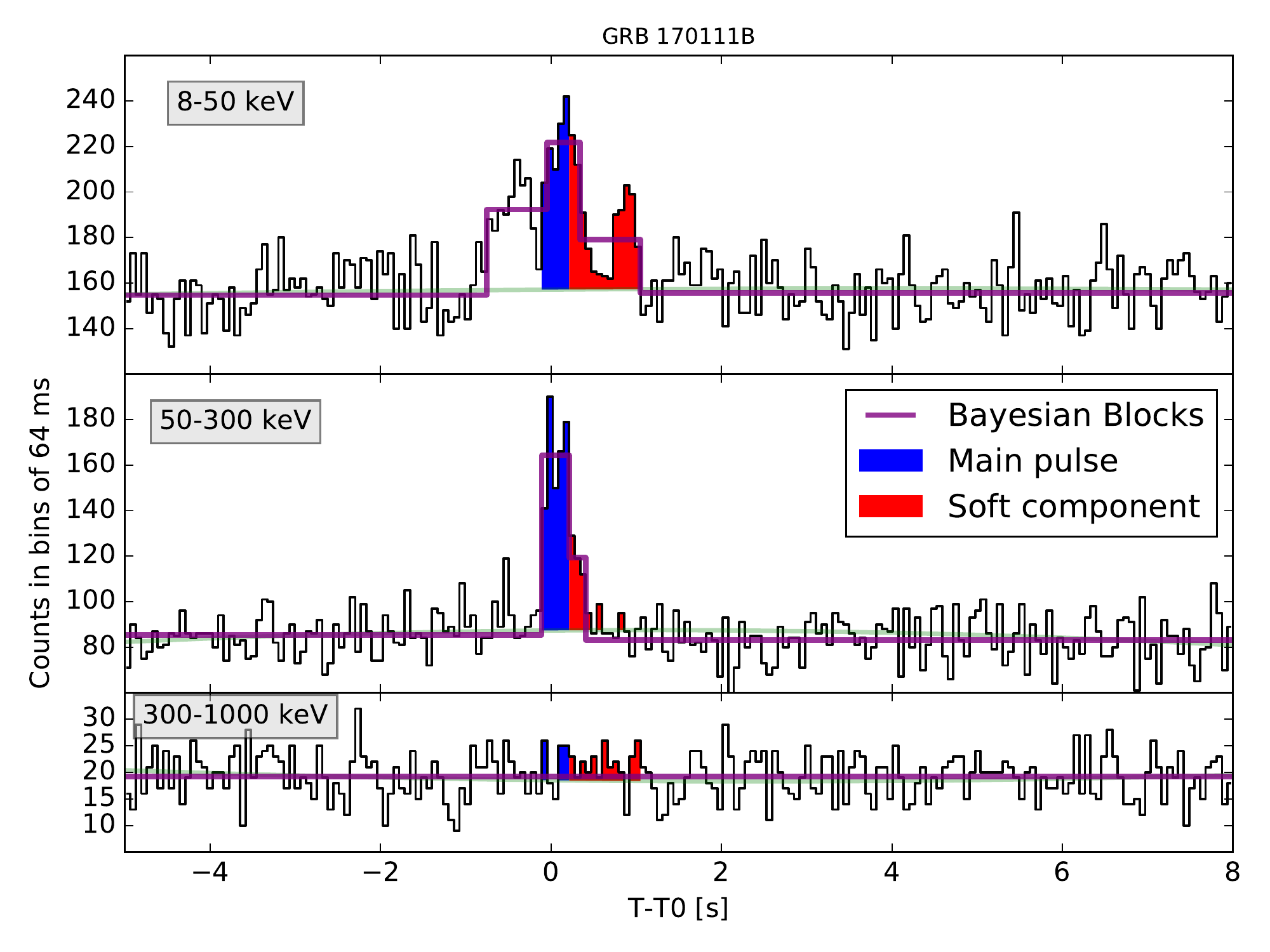}
\end{tabular}
\caption{\label{fig:GRB170817AthreeEBins} Left: composite light curves using NaI(Tl) detectors in the 8--50 keV (top panel), 50--300 keV (middle panel), and  300--1000 keV (bottom panel) energy ranges for \grb. Right: Same, but for GRB 170111B. Blue marks the main pulse,  red is the soft tail, and the procedure to find their span is given in Section \ref{sec:BayesBlockAnal}. The observed pretrigger soft emission of GRB 170111B is later discussed in Section~\ref{sec:Disc}.}
\end{figure}

As the main pulse of \grb was dominant over the 50--300 keV energy range\footnote{GRBs routinely observed by GBM typically have the highest count rate in this spectral range.} and the soft episode was strongest below 50 keV, we consider three energy ranges for our analysis: 8--50, 50--300, and 300--1000~keV. We run the Bayesian Block algorithm over the light curves within these energy ranges (see Figure \ref{fig:GRB170817AthreeEBins} for an example), designating the main pulse as the brightest nonbackground Bayesian Block interval in the 50--300~keV light curve. As we are looking at triggered GRBs, at least one Bayesian Block will necessarily exist in this energy range. If there are non-background blocks before the peak, those would be assigned to the main pulse as well. If the Bayesian Block model in the 8--50~keV lightcurve identifies emission that lasts longer than the main emission episode in the 50--300~keV range, then we attribute this as possible soft emission. In some cases, the Bayesian Block algorithm failed, because it did not detect the GRB over the 8--50~keV channel. 

The calculation of the hardness ratio (ratio of photon counts in 50--300 and  8--50~keV ranges) for the main and tail emissions obtained by the Bayesian Block method is also invoked as an additional criterion in our analysis. In order to identify sGRBs with characteristics similar to \grb, we required the main peak hardness ratio to be greater than 1 (HR$_{\rm \,main} > 1$) and the soft tail hardness ratio to be smaller than 1 (HR$_{\rm \, soft} < 1$). In this way, we identify 327 candidates out of our initial sample of 558 GRBs. 
However, this sample is contaminated by bursts showing a clear hard-to-soft evolution and by several bursts displaying a relatively high flux in the 50--300~keV range, comparable to the observed flux in the 8--50~keV range. Thus it was necessary to further inspect it, using an additional manual selection process that obeyed the following criteria:

\begin{itemize}
\item A significantly luminous initial peak, which is brighter over 50--300~keV than the 8--50~keV energy range.
\item A weak tail, which is bright over the 8--50~keV energy range and disappears at higher energies.
\item A discernible change of the light curve, giving some indication of two emission mechanisms.
This criteria was used to avoid continuous hard-to-soft spectral evolution, which
is extremely common in GRB pulses.
\end{itemize}
The manual check resulted in 17 candidates (including \grb and  GRB 150101B, which were easily identified by this method).
Visually inspecting the sample of GRBs, not selected by the constraint on the HR of the main and tail emission, we found no likely additional candidates, confirming the Bayesian Blocks method as a good prefilter.
These 17 candidates still do not represent the final sample, due to two important, additional selection criteria that need to be fulfilled:  
\begin{itemize}
\item The localization of the main and soft emission episodes must coincide.
\item The spectral characteristics of the soft tail must be similar to that of \grb.
\end{itemize}
The fulfillment of these criteria is checked and presented in the next two sections.

\subsection{Localization Analysis}
\label{sec:localization}
The Bayesian Block method identifies temporally coincident excesses; however, we cannot determine a connection through temporal coincidence alone. 
The GBM background is dominated by real signals from galactic transients, solar flares, magnetospheric particle precipitation events, etc.
 Therefore, we must also verify that both components arise from the same location. 
GBM can perform degree-level accuracy localizations of signals using a standard technique originally developed for BATSE and adapted for use to localize all GBM triggers.
Provided a manual selection identifying the signal and a background model, the localization algorithm
compares the relative count rates observed in the detectors and estimates the most likely arrival direction by performing a chi-squared minimization over an 1$^\circ$ grid on the sky and three spectral templates. The spectral templates are Band functions \citep{Band+93}, representing spectrally hard, normal, and soft GRBs as observed by GBM. The main peak of each GRB was localized utilizing all three templates, while the soft tail was localized using only the softer spectrum. The localization also include an additional systematic component determined as a core-plus-tail model, with 90~\%\ of GBM GRBs having a 3.7$^\circ$ systematic error and a tail of greater than 10$^\circ$ \citep{2015ApJS..216...32C}.

Using by-eye selections of source windows, independent of those found through the Bayesian Block analysis, we separately localized the main peak and the subsequent softer emission for all GRB candidates in our sample. For 15 out of the 17 candidates, we derived consistent localizations. Two candidates were withdrawn at this step of the analysis as the soft tail could not be localized.   

\floattable
\begin{deluxetable}{lrccccrcccc}
\rotate
\tabletypesize{\scriptsize}
\tablecaption{\label{tab:candlist} Spectral analysis results for 15 candidate GRBs. GRBs marked with stars were rejected (for details see Sec.~\ref{sec:spec_analysis}).}
\tablehead{ \\
&  &  &  &  & &  & & Energy Flux (EF) & Fluence ($\mathscr{F}$) & \\
GRB Name& \multicolumn{1}{c}{Time Int} & Model & Epeak & Index & $kT$ & C-Stat/DOF & Phot. Flux & $\times 10^{-7}$ & $\times 10^{-7}$ & $\mathscr{F}_{T}$/$\mathscr{F}_{P}$ \\
& \multicolumn{1}{c}{(s)} &  & (keV) &  & (keV) &  & (ph cm$^{-2}$ s$^{-1}$) & (erg cm$^{-2}$ s$^{-1}$) &  (erg cm$^{-2}$)& }
\startdata
&-0.128:0.256&Comp&1473 $\pm$ 275 & -0.75 $\pm$ 0.08& - & 366.8/359 & 10.8 $\pm$ 0.6 & 40.14 $\pm$ 0.20 & 15.4 $\pm$ 0.8 & \\ \cline{2-10}
GRB 081209A & 0.384:0.768 &PL& - & -2.13 $\pm$ 0.35 & - & 368.8/360 & - & - & -  &  \\
& 0.384:0.768 & \bf{BB} & - & - & 5.9 $\pm$ 1.4 & 369.8/360& 1.8 $\pm$ 0.5 & 0.60 $\pm$ 0.17 & 0.23 $\pm$ 0.07 & 0.015  \\
\hline
 &0.000:0.128 &Comp&811 $\pm$ 31 & -0.19 $\pm$ 0.04 & - & 533.7/486&  100.3 $\pm$ 1.6 &  468.0 $\pm$ 8.3 & 59.9 $\pm$ 1.1 & \\ \cline{2-10}
GRB 090228A* &  0.256:0.512 &\bf{Comp}&92 $\pm$ 13&-0.91 $\pm$ 0.22& - & 512.7/486 & 9.7 $\pm$ 0.6 & 7.48 $\pm$ 0.67 & 1.91 $\pm$ 0.10 & 0.032\\
&0.256:0.512&BB& - & - & 17.0 $\pm$ 0.8&571.7/487& - & - & -  & \\
\hline
&-0.064:0.384 &Comp&927 $\pm$ 177 & -0.54 $\pm$ 0.10 & - & 697.7/606 &  8.6 $\pm$ 0.3 &  33.61 $\pm$ 1.70 & 15.1 $\pm$ 0.76 & \\ \cline{2-10}
GRB 100328A &  1.024:1.344&Comp&34 $\pm$ 7&-0.12 $\pm$ 1.30& - & 652.9/606 & - & - & - & \\
&1.024:1.344&\bf{BB}& - & - & 8.2 $\pm$1.2&653.4/607& 2.7 $\pm$ 0.4 & 1.14 $\pm$ 0.20 & 0.36 $\pm$ 0.06 & 0.024 \\
\hline
   &-0.384:0.256 &Comp& 248 $\pm$ 71 & 0.36 $\pm$ 0.79 & - & 390.5/360 &  1.8 $\pm$ 0.3 &  4.21 $\pm$ 1.10 & 2.69 $\pm$ 0.70 & \\ \cline{2-10}
GRB 100719C* &  0.256:0.640&\bf{PL}& - &-1.57 $\pm$ 0.19& - & 369.9/361 &2.07 $\pm$ 0.52  &2.95 $\pm$ 0.90 & 1.13 $\pm$ 0.35 & 0.42\\
 &0.256:0.640& BB & - & - & 25.3 $\pm$ 6.2&375.1/361& - & - & - & \\
 \hline
 &-0.256:0.256&Comp&341 $\pm$ 320 & -1.04 $\pm$ 0.39 & - & 487.0/486 & 2.5 $\pm$ 0.4 & 4.04 $\pm$ 0.15 & 2.07 $\pm$ 0.77 & \\ \cline{2-10}
GRB 101224A & 1.280:2.048&PL& - & -2.09 $\pm$ 0.39 & - & 565.3/487 &  - & - &  - &   \\
&1.280:2.048&\bf{BB}& - & - & 8.2 $\pm$ 2.2 &567.2/487& 0.85 $\pm$ 0.26 & 0.36 $\pm$ 0.12 & 0.28 $\pm$ 0.09 & 0.14\\
\hline
 &0.000:0.128&Comp&328 $\pm$ 67&-0.34 $\pm$ 0.26& - & 326.2/363 &  10.7 $\pm$ 1.0 & 25.57 $\pm$ 3.50 & 3.27 $\pm$ 0.45 & \\ \cline{2-10}
GRB 110717A & 0.384:0.768 &PL& - &-2.45 $\pm$ 0.66& - & 418.4/364 & - & - & - & \\
&0.384:0.768&\bf{BB}& - & - & 7.1 $\pm$ 2.2 & 417.9/364 & 1.3 $\pm$ 0.4 & 0.49 $\pm$ 0.18 & 0.19 $\pm$ 0.07 &0.06\\
\hline
&-0.128:0.256&Comp&144 $\pm$ 18&0.53 $\pm$ 0.60& - & 395.7/365& 2.8 $\pm$ 0.4 & 4.11 $\pm$ 0.53 & 1.58 $\pm$ 0.20 & \\  \cline{2-10}
GRB 111024C & 0.768:1.409&PL& - &-2.02 $\pm$ 0.49& - & 408.7/366 & - & - & - &  \\
&0.768:1.409&\bf{BB}& - & - & 8.7 $\pm$ 2.6 & 406.6/366& 0.9 $\pm$ 0.3 & 0.39 $\pm$ 0.15 & 0.25 $\pm$ 0.10 & 0.16 \\
\hline
& -0.128: 0.384 &Comp&133 $\pm$ 20&0.66 $\pm$ 0.68 & - & 542.9/489 & 2.7 $\pm$ 0.4 & 3.72 $\pm$ 0.53 & 1.90 $\pm$ 0.27 & \\ \cline{2-10}
GRB 120302B  &  0.640:1.792&PL&- &-2.38 $\pm$ 0.43& - & 556.7/490& - & - & - &  \\
& 0.640:1.792 & \bf{BB} &  - & - & 6.9 $\pm$ 1.4&552.2/490 & 1.1 $\pm$ 0.3 & 0.40 $\pm$ 0.09 & 0.46 $\pm$ 0.10 & 0.24\\
\hline
& -0.128:0.384 & Comp & 526 $\pm$ 114 & -0.21 $\pm$ 0.25 & - & 523.9/464& 3.8 $\pm$ 0.3 & 13.66 $\pm$ 1.40 & 6.99 $\pm$ 0.72 & \\ \cline{2-10}
GRB 120915A & 0.640:1.280 &PL & - & -1.89 $\pm$ 0.45 & - & 475.6/465& - & - & - & \\
& 0.640:1.280 & \bf{BB} & - & - & 10.2 $\pm$ 2.7& 471.6/465 & 0.8 $\pm$ 0.2 & 0.38 $\pm$ 0.13 & 0.24 $\pm$ 0.08 & 0.034\\
\hline
& -0.512:1.024&Comp&91 $\pm$ 20&-0.80 $\pm$ 0.35 & - & 448.5/363 & 2.7 $\pm$ 0.3 & 2.12 $\pm$ 0.30 & 3.26 $\pm$ 0.46 & \\ \cline{2-10}
GRB 130502A & 2.048:3.072&Comp&54 $\pm$ 21 &-1.26 $\pm$ 0.59& - & 393.9/363& - & - & - &  \\
& 2.048:3.072 & \bf{BB} &  - & - & 10.1 $\pm$ 1.2 & 403.5/364 & 1.8 $\pm$ 0.3 & 0.88 $\pm$ 0.14 & 0.90 $\pm$ 0.14 & 0.28\\
\hline
& -0.064:0.128&Comp&280 $\pm$ 58&-0.78 $\pm$ 0.16 & - & 815.0/727& 8.7 $\pm$ 0.6 & 14.85 $\pm$ 1.70  & 2.85 $\pm$ 0.33 & \\ \cline{2-10}
GRB 140511A & 0.128:0.384&Comp&29 $\pm$ 42 &-1.79 $\pm$ 0.63& - & 727.7/727& - & - & - &  \\
& 0.128:0.384 & \bf{BB} & - & - & 6.7 $\pm$ 1.0 &728.7/728 & 2.6 $\pm$ 0.4 & 0.94 $\pm$ 0.15& 0.24 $\pm$ 0.04 & 0.084 \\
\hline
& -0.016:0.000&Comp&524 $\pm$ 176&-0.80 $\pm$ 0.20 & - & 638.2/885 & 28.4 $\pm$ 2.6 & 70.58 $\pm$ 8.40 & 1.13 $\pm$ 0.13 & \\ \cline{2-10}
GRB 150101B & 0.000:0.064&PL& -  &-2.42 $\pm$ 0.21& - & 723.1/886& - & - & - &  \\
& 0.000:0.064 & \bf{BB} & - & - & 6.0 $\pm$ 0.6 &713.3/886 & 9.2 $\pm$ 1.1 & 3.07 $\pm$ 0.37& 0.20 $\pm$ 0.02 &0.18\\
\hline
 & -0.768:-0.192 & Comp & 49 $\pm$ 7 & -0.08 $\pm$ 0.70 & & 708.5/633& - & - & - &   \\
& -0.768:-0.192 & \bf{BB} &  - & - & 10.8 $\pm$  1.0 & 709.8/634 & 2.5 $\pm$ 0.3 & 1.31 $\pm$ 0.15 & 0.75 $\pm$ 0.09 & 0.20 \\ \cline{2-10}
GRB 170111B  & -0.128:0.384 & Comp & 154 $\pm$ 22 & -0.62 $\pm$  0.19 & - & 697.0/633 & 6.3 $\pm$ 0.4 & 7.42 $\pm$ 0.70 & 3.80 $\pm$ 0.36 & 0.25 \\ \cline{2-10}
  & 0.768:0.960 & Comp & 35 $\pm$ 13 & -1.12 $\pm$ 1.03 & - & 608.5/607& - & - & - &   \\
& 0.768:0.960 & \bf{BB} & - & - & 8.1 $\pm$  1.0 & 611.8/608 & 2.6 $\pm$ 0.5 & 1.08 $\pm$ 0.22 & 0.21 $\pm$ 0.04 & 0.06\\
\hline
 & -0.512:0.512 & Comp & 197 $\pm$ 89 & -0.84 $\pm$  0.39 & - & 527.3/506 & 1.6 $\pm$ 0.2 & 2.11 $\pm$ 0.56 & 2.16 $\pm$ 0.57 & \\ \cline{2-10}
GRB 170817A  & 0.512:2.048 & PL & - & -1.99 $\pm$ 0.26 & - & 639.0/507& - & - & - & \\
 & 0.512:2.048 & \bf{BB} & - & - & 11.2 $\pm$  1.5 & 622.4/507 & 0.9 $\pm$ 0.2 & 0.49 $\pm$ 0.09 & 0.75 $\pm$ 0.14 &0.35\\
\hline
& -0.032:0.032 & Comp & 639 $\pm$ 220 & -0.61 $\pm$  0.22 & - & 697.9/717 & 10.8 $\pm$ 1.0 & 34.29 $\pm$ 3.80 & 2.19 $\pm$ 0.24 & \\ \cline{2-10}
GRB 180511A  & 0.032:0.128 & PL &  - & -1.97 $\pm$ 0.45 & - & 671.6/718& - & - & - &   \\
& 0.032:0.128 & \bf{BB} & - & - & 11.1 $\pm$  3.0 & 667.4/718 & 1.9 $\pm$ 0.6 & 1.00 $\pm$ 0.33 & 0.10 $\pm$ 0.03 & 0.046\\
\enddata
\label{tab:candlist}
\end{deluxetable}

\floattable
\begin{deluxetable}{llcrrrrrrrll}[t]
\rotate
\tabletypesize{\footnotesize}
\tablecaption{\label{tab:candlist_temp} Standard \Fermi-GBM burst catalog parameters of the final sample of 13 candidate GRBs, which is including the reference \grb.}
\setlength{\tabcolsep}{+3pt} 
\tablehead{ & & &\multicolumn{2}{c}{Durations}& \multicolumn{3}{c}{Localization} &\colhead{Total Fluence} & \colhead{Peak Flux} & & \\ GRB Name & Trigger ID\tablenotemark{a} & \colhead{Time} & \colhead{\tn} & \colhead{\tf} & \colhead{R.A.} & \colhead{Decl.} & \colhead{Error}  &  \colhead{(erg cm$^{-2}$)}  & \colhead{(64 ms)}  &\colhead{Detect.\tablenotemark{d}} & \colhead{References} \\
& & (UTC) & \multicolumn{1}{c}{(s)} & \multicolumn{1}{c}{(s)} & (deg.) & (deg.) & (deg.) & \multicolumn{1}{c}{$\times 10^{-7}$} & (ph cm$^{-2}$ s$^{-1}$) & &  }
\startdata
 GRB~081209A\tablenotemark{b} & bn081209981 & 23:41:56.39 &  0.192 $\pm$ 0.143 &  0.128 $\pm$ 0.143  & 45.3 & 63.5 & 4.9 & 14.66 $\pm$ 1.49 & 25.4 $\pm$ 1.2 & KW,S\tablenotemark{e},A & \cite{GCN8646,GCN8647}  \\
 GRB~100328A\tablenotemark{b}& bn100328141 & 03:22:44.60 &  0.384 $\pm $0.143 & 0.192 $\pm$ 0.091  & 155.9 & 47.0 & 4.8 & 10.01 $\pm$ 0.24 & 13.4 $\pm$ 0.8 &  & \cite{Abadie+12oldligo} \\
 GRB 101224A & bn101224227 & 05:27:13.86 & 1.728 $\pm$ 1.68 & 0.192 $\pm$ 0.286  & 285.9 & 45.7 & 0.1\tablenotemark{f} & 1.92 $\pm$ 0.27 & 6.7 $\pm$ 1.0 & S & \cite{GCN11484,GCN11491}; \\
 & & &  &  & & &  &  & &  & \cite{,GCN11492,GCN11596} \\
 GRB~110717A\tablenotemark{b} &bn110717180 & 04:19:50.66 & 0.112 $\pm$ 0.072 & 0.032 $\pm$ 0.023  & 308.5 & -7.9 & 7.5 & 2.51 $\pm$ 0.12 & 18.5 $\pm$ 1.8 & KW, IA & \Fermi-GBM Only \\ 
 GRB~111024C\tablenotemark{b} & bn111024896 & 21:30:02.24 &  0.960 $\pm$ 1.032 & 0.256 $\pm$ 0.143  & 91.2 & -1.8 & 13.2 & 3.80 $\pm$ 0.16 & 7.4 $\pm$ 1.2 &IA & \Fermi-GBM Only \\
 GRB~120302B\tablenotemark{b} & bn120302722 & 17:19:59.08 &  1.600 $\pm$ 0.779 & 0.512 $\pm$ 0.466  & 24.1 & 9.7 & 13.9 & 1.19 $\pm$ 0.16 & 6.2 $\pm$ 1.5 & & \Fermi-GBM Only \\ 
 GRB~120915A\tablenotemark{c} & bn120915000 & 00:00:41.64 & 0.576 $\pm$ 1.318 & 0.320 $\pm$ 0.091  & 209.4 & 67.3 & 5.9 & 5.06 $\pm$ 0.26 & 6.0 $\pm$ 0.9 & IA, SW & \Fermi-GBM Only \\ 
 GRB~130502A & bn130502743  & 17:50:30.74 & 3.328 $\pm$ 2.064 & 2.304 $\pm$ 0.572 & 138.6 & -0.1 & 0.0\tablenotemark{f} & 6.27 $\pm$ 0.35 & 6.6 $\pm$ 1.4 & S, OT & \cite{GCN14527,GCN14531}; \\
 & & & &  & & &  &  & &  & \cite{GCN14533}; \\
 & & & &  & & &  &  & &  & \cite{GCN14535,GCN14543} \\ 
 GRB~140511A\tablenotemark{c} & bn140511095 & 02:17:11.56 & 1.408 $\pm$ 0.889 & 0.256 $\pm$ 0.181  & 329.8 & -30.1 & 8.8 & 3.71 $\pm$ 0.32 & 9.4 $\pm$ 1.0 & & \Fermi-GBM Only \\
 GRB~150101B & bn150101641  & 15:23:34.47 & 0.08 $\pm$ 0.928 & 0.016 $\pm$ 0.023  & 188.0 & -11.0 & 0.0\tablenotemark{f} & 2.38 $\pm$ 0.15 & 10.5 $\pm$ 1.3 &  S, IA, C, & \cite{GRB150101B_Troja,2018ApJ...863L..34B}; \\
 & & & &  & & &  &  & & X, z &\cite{GRB150101B_fong} \\
 GRB~170111B\tablenotemark{c} & bn170111815 & 19:34:01.39 & 3.072 $\pm$ 1.318 & 0.32 $\pm$ 0.091  & 270.9 & 63.7 & 6.7 & 5.96 $\pm$ 0.12 & 7.6 $\pm$ 1.0 & & \Fermi-GBM Only \\
 GRB~170817A  & bn170817529 & 12:41:06.47 & 2.048 $\pm$ 0.466 & 1.28 $\pm$ 0.405  & 197.5 & -23.4 & 0.0\tablenotemark{f} & 2.79 $\pm$ 0.17 & 3.7 $\pm$ 0.9 & L, z, C, & \cite{2017ApJ...848L..12A} \\ 
 & & & &  & & &  &  & & IA, HST  & \\ 
 & & & &  & & &  &  & & and more & \\
 GRB~180511A\tablenotemark{c} & bn180511364 & 08:43:35.79 &  0.128 $\pm$ 1.207 & 0.032 $\pm$ 0.045  & 250.4 & -8.2 & 15.1 & 1.53 $\pm$ 0.21 & 9.2 $\pm$ 1.0 & IA & \Fermi-GBM Only \\
\enddata
\tablenotetext{a}{Burst candidates, whose ID or burst number ("bn'') has the format bnYYMMDDXXX. Y, M and D are the year, month and day of the trigger respectively. The fraction of the day is given by the last three numbers in the ID ("XXX'').}
\tablenotetext{b}{GRBs which occurred during the time span covered by the  2nd \Fermi-GBM burst catalog \citep{2014ApJS..211...13V} and weren't reported via GCN by other instruments, were designated being the “A”,“B” or “C” burst of that day, depending on their chronological order.}
\tablenotetext{c}{Bursts for which no follow-up was initiated via GCN, and are named here for the first time. Lettering was determined by cross-checking these bursts against the IPN master burst list (\url{http://www.ssl.berkeley.edu/ipn3/masterli.txt}).}
\tablenotetext{d}{KW: Konus-Wind, A: AGILE-MCAL, IA: \INTEGRAL\ SPI-ACS, S: \Swift, L: LIGO, C = Chandra, X: XMM-Newton, SW: Suzaku-WAM, HST = Hubble Space Telescope, OT = Optical Transient, z = VLT/Redshift.}
\tablenotetext{e}{Outside of the coded field of BAT}
\tablenotetext{f}{The RA, declination and their associated errors are from other missions, resulting in a more accurate localization. Errors which are ``0.0" are extremely small and therefore effectively zero.}
\label{tab:GRBcanCatPar}
\end{deluxetable}

\subsection{Spectral Analysis}
\label{sec:spec_analysis}
The spectral analysis of the 15 remaining candidate bursts was performed using 
RMfit\footnote{The spectral analysis package RMfit was originally developed for time-resolved analysis of BATSE GRB data but has been adapted for GBM and other instruments with suitable FITS data formats. The software is available at the Fermi Science Support Center:~\href{http://fermi.gsfc.nasa.gov/ssc/data/analysis/user/}{http://fermi.gsfc.nasa.gov/ssc/data/analysis/user/}. A tutorial is also available at \href{http://fermi.gsfc.nasa.gov/ssc/data/analysis/scitools/rmfit_tutorial.html}{http://fermi.gsfc.nasa.gov/ssc/data/scitools/rmfit\_tutorial.html}.}.
The assumed photon model (i.e.  comptonized (Comp), power-law (PL), BB, etc.) is folded through the detector response to produce a model counts spectrum, which is then compared to the observed counts spectrum. A nonlinear least squares minimization method is then performed in counts spectrum space to produce a spectral fit. Since the detection efficiency of the NaI(Tl) detectors decrease for large incidence angles, we chose detectors with source angles $\leq 60$ \citep{2009ExA....24...47B}, as is standard GBM procedure. More on this and other important steps in the spectral analysis of the bursts presented in this paper can be found in~\citet{Gruber2014}.

The results of the spectral analysis are summarized for  15 GRBs in Table \ref{tab:candlist}. These include the results for \grb and GRB 150101B. For each of the GRBs listed in the table, the individual rows show the results of the spectral analysis over the time intervals of the main and tail emission. The exception is GRB 170111B, where we show spectral analysis of the soft pretrigger emission in addition to the other results. 
For the main emission episode, only the values for $E_{\rm peak}$ and the index are shown (first row for each GRB), derived from fits with the Comptonized (Comp) model, which was in all cases statistically preferred compared to fits with the PL or BB model. This allows us to determine if the main peak is spectrally harder when compared to the tail emission. 
The spectral analysis results for the tail emission, shown for each GRB in the second and third rows, compare the Comp or PL fit results with the one derived from a fit with a BB model. 
We used the difference of the Castor C-Statistic (C-Stat, see \citet{Goldstein+17170817} for a comparison of C-Stat values for different models based on simulations) per degree of freedom (DOF) in order to assess which model fits the data best, or equally well (model name highlighted in bold face).  
In cases where the fit parameters of the Comp were not constrained, results using the simpler PL model are presented. The last four columns list the energy flux (EF), fluence ($\mathscr{F}$) and fluence ratio $\mathscr{F}_{T}$/$\mathscr{F}_{P}$ for the selected tail model and peak (main) model  mission respectively. In the case of GRB 170111B, the fluence ratio is calculated individually for the pretrigger and tail emission, in addition to the sum of both soft emission periods, listed in the middle row.

We considered GRBs as possible candidates, in which the tail emission was fitted  equally well or better  by a BB  compared to the  PL or Comp models, giving an indication of the tail emission having thermal characteristics. 
Since only one GRB/GW event has been identified up to now and no commonly agreed theoretical guidance is available, we set a tentative limit to the BB temperature $kT$ for a candidate GRB being similar to \grb. In this study, we set this BB temperature limit at about twice the value of \grb, i.e. $kT\lesssim$ 20~keV. Using this approach, we exclude candidate GRB 100719C (see row 4 of Table \ref{tab:candlist}). Moreover, we also excluded GRB 090228A (see row 2 of Table \ref{tab:candlist}), because it is showing a  lower value of C-Stat for the Comp model than for the BB  model. This hints to the possibility that the tail emission is not similar to the BB tail of \grb. 
We have not removed more GRBs from the candidate list, so as not to restrict the full range of candidates, some of them being clear candidates and others being more doubtful, 
 e.g.\ we kept GRB 130502A in the list, although it has a  slightly larger value of C-Stat for the BB model compared to the Comp model.
Additionally, we will also not provide a ranking of our candidates at this stage until the detection of new GRB/GW events have consolidated the behaviour of these phenomenon. This final step leaves us with 13 candidates, including \grb and GRB 150101B. The light curves for this final list of candidates are shown in Figures \ref{fig:lc4} and \ref{fig:lc5}. The final limits of the main pulse and soft tail were determined by considering the temporal bins with sufficient flux above the background for spectral analysis. These are not necessarily the same as the Bayesian Block intervals.

In order to put the candidate list into context with the whole GBM GRB database, we list the standard burst catalog parameters, like the corresponding trigger ID, the trigger time in UTC,  \tn , \tf, the localization, the total fluence during the \tn\, time interval and the 64 ms peak flux in Table \ref{tab:GRBcanCatPar}. The second last column in this table shows which of the GRBs were detected or observed by other instruments and the corresponding references are listed in the last column.

\section{Properties of the Final Candidates}
\label{sec:Can_Samp}

In order to further inspect the credibility of our candidates and reveal any new individual or collective characteristics for the GRBs in our sample, their properties were studied. This was done by employing different analytical tools, like the targeted search, anisotropy tests of the spatial distribution, determination of the spectral hardness distribution, pulse fitting, and variability checks of individual bursts, and correlation analysis of the parameters derived from the spectral analysis.

\subsection{Targeted Search Characterization}
\label{sec:TargSearch}

\begin{figure}[b!]
\centering
\begin{tabular}{cc}
\includegraphics[width=0.50\textwidth]{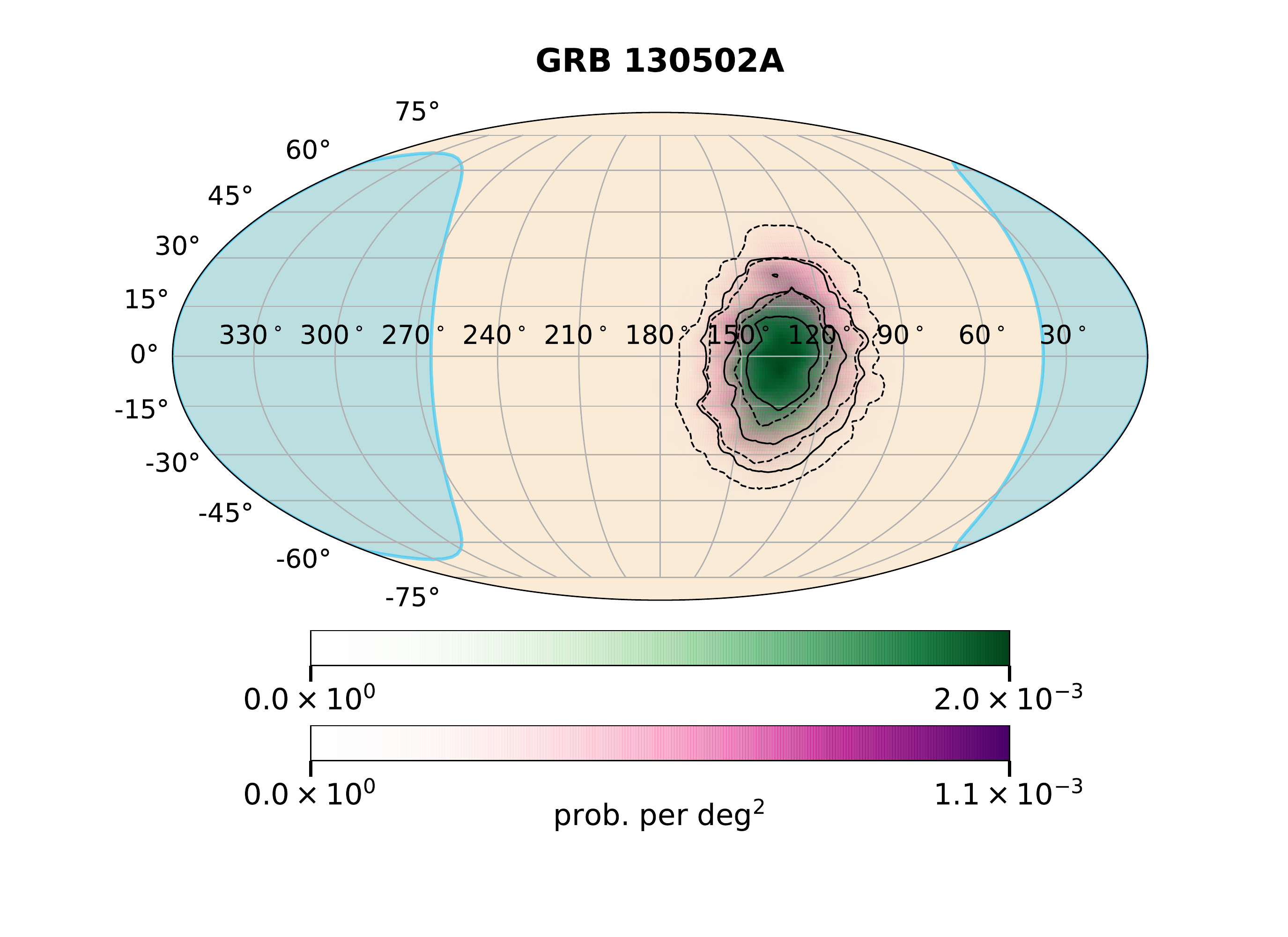} &
\includegraphics[width=0.50\textwidth]{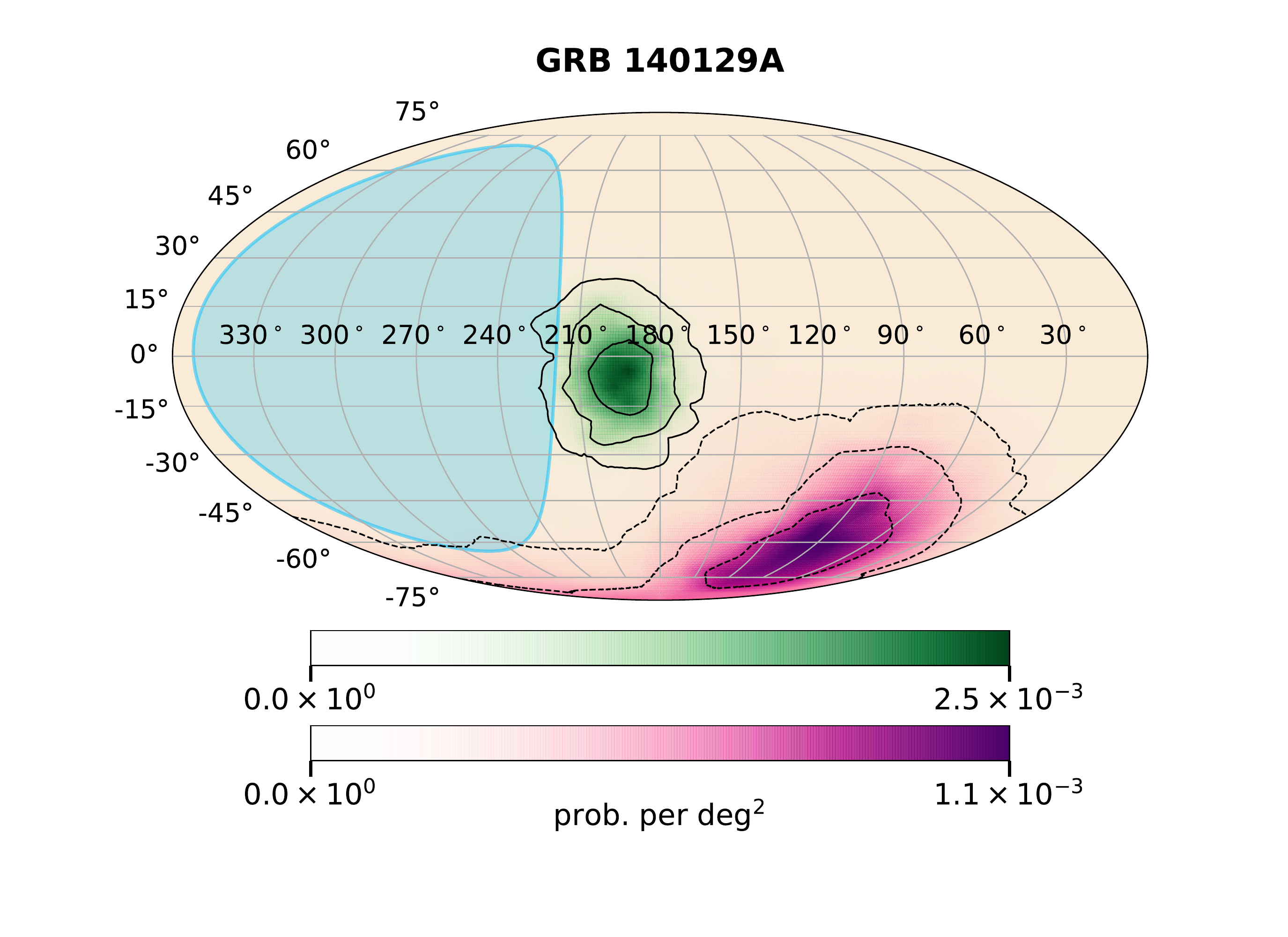} \\
\end{tabular}
\caption{\label{fig:Targ_loc} Two examples of the Targeted Search localization contours: GRB 130502A and GRB 140129A. The localization of the main peak is shown in green and the candidate soft tail displayed in purple. Contours are plotted at the 1$\sigma$, 2$\sigma$, and 3$\sigma$ containment regions. The region of sky occulted by the Earth is shown in blue. For GRB 130502A, the spatial agreement between the main peak and soft tail is 0.91, while for GRB 140129A the spatial agreement is 7.65E-4.}
\end{figure}

The six candidate GRBs detected after 2012 and GRB 140129A (trigger ID: bn140129499), a withdrawn candidate from the localization analysis (Section2.4) were further investigated by applying the GBM Targeted Search \citep{Blackburn2015} to verify the results from the Bayesian Block and spectral analysis as well as provide additional characterization of the spectral components. The Targeted Search was developed to find gamma-ray transients in the GBM data that lie below the on-board triggering threshold. It is used to follow up GW triggers from LIGO/Virgo (Burns et al. 2018) and has been demonstrated to recover Swift short GRBs within the GBM field of view (Kocevski et al. 2018). Due to improvements detailed in Goldstein et al. (2016), particularly that of the background estimation method, the Targeted Search can yield improved localizations over those found by techniques relying on manual selections of source and background. This search was utilized to confirm whether both spectral components (main peak and soft tail), were consistent with the same location. Furthermore, we attempted to localize the candidate soft tail of GRB 140129A with the Targeted Search due to its sensitivity to weak events, a task  that we could not accomplish manually. The Targeted Search is based on the analysis of CTTE data, which are available only after 2012 November. Consequently, this method could not be applied to the earlier seven candidate events in our sample.

\begin{table}[b!]
\begin{center}
\caption{Summary of Targeted Search results for each GRB in our sample.}
\begin{tabular}{c | cccc | ccccc | c} 
\hline \hline
& \multicolumn{4}{c|}{Main Peak} & \multicolumn{5}{c|}{Blackbody} &  Spatial \\
GRB & Start* & Duration & RA & Dec & Start* & Duration & R.A. & Decl. & S/N & Probability \\
  & (s) & (s) & (deg) & (deg) & (s) & (s) & (deg) & (deg) & & \\
\hline
GRB 130502A & -0.054 & 0.512 & 136.9 & 0.1 & 1.977 &  0.128 & 134.12 & 9.80 & 8.54 & 0.912 \\
 \hline
GRB 140129A & -0.033 & 0.064 & 197.5 & -6.4 & 2.783 & 0.064 & 86.83 & -63.98 & 5.36 & 7.652E-4 \\
 \hline
GRB 140511A & -0.049 & 0.128 & 312.7 & -28.8 & 0.127 & 0.032 & 318.22 & -27.49 &  4.59 & 0.792\\
 \hline
 GRB 150101B & -0.017 & 0.016 & 188.2 & -4.6  & -0.001 & 0.064 & 163.6 & -23.1  &  10.49 & 0.846 \\
 \hline
GRB 170111B & -0.128 & 0.512 & 254.3 & 59.1 & 0.751 & 0.256 & 224.69 & 62.20 & 7.77 & 0.893\\
 \hline
 GRB 170817A & -0.273 & 0.512 & 178.1 & -33.1 & -0.113 & 2.048 & 189.73 & -17.31 & 9.53 & 0.814\\
 \hline
GRB 180511A & -0.031 & 0.032 & 252.3 & 0.7 & 0.001 & 0.256 & 266.98 & 6.77 & 5.03 & 0.850\\
 \hline
\end{tabular}
\label{tab:search_results}
\end{center}
\end{table}

\begin{figure}[t!]
\centering
\includegraphics[width=0.6\textwidth]{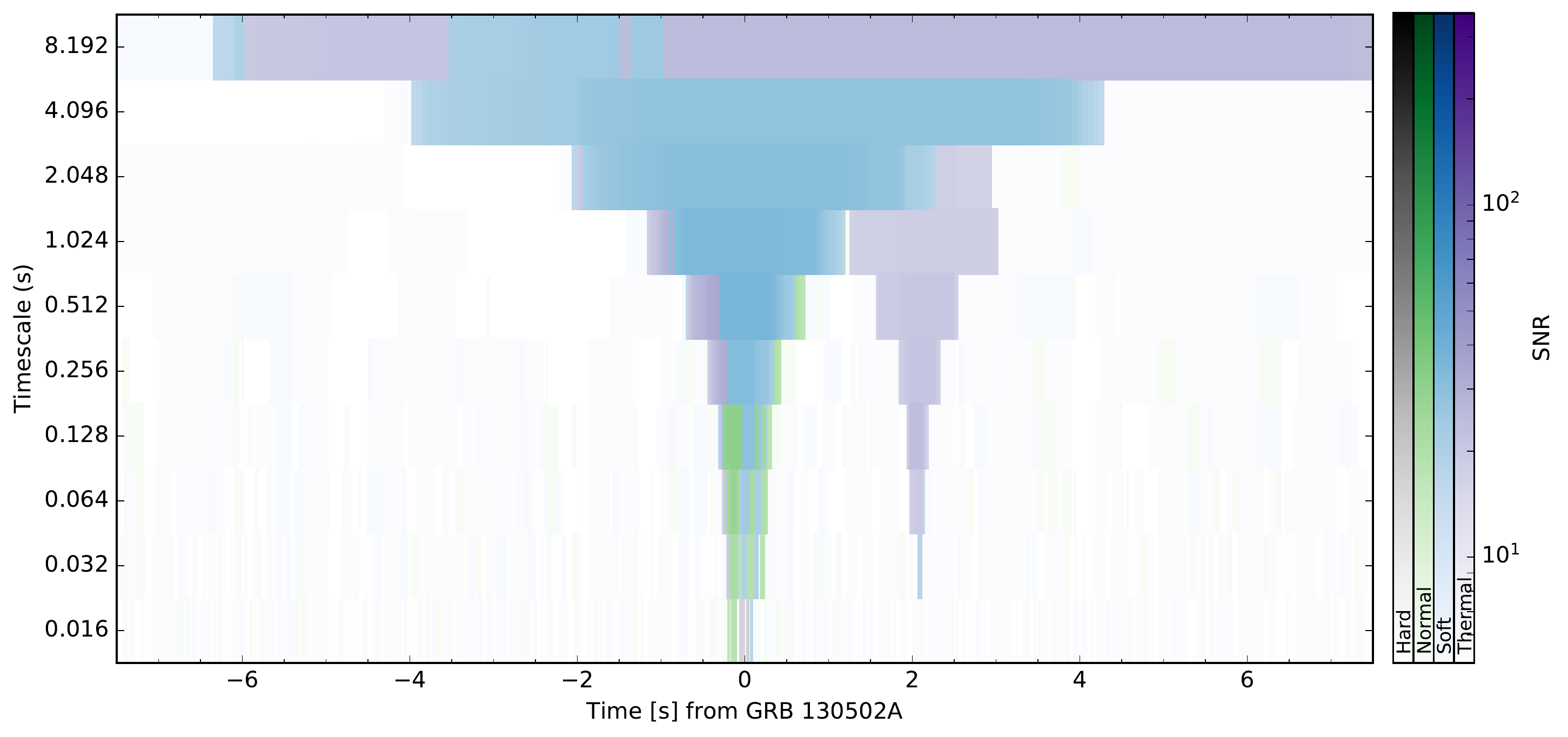} \\
\includegraphics[width=0.6\textwidth]{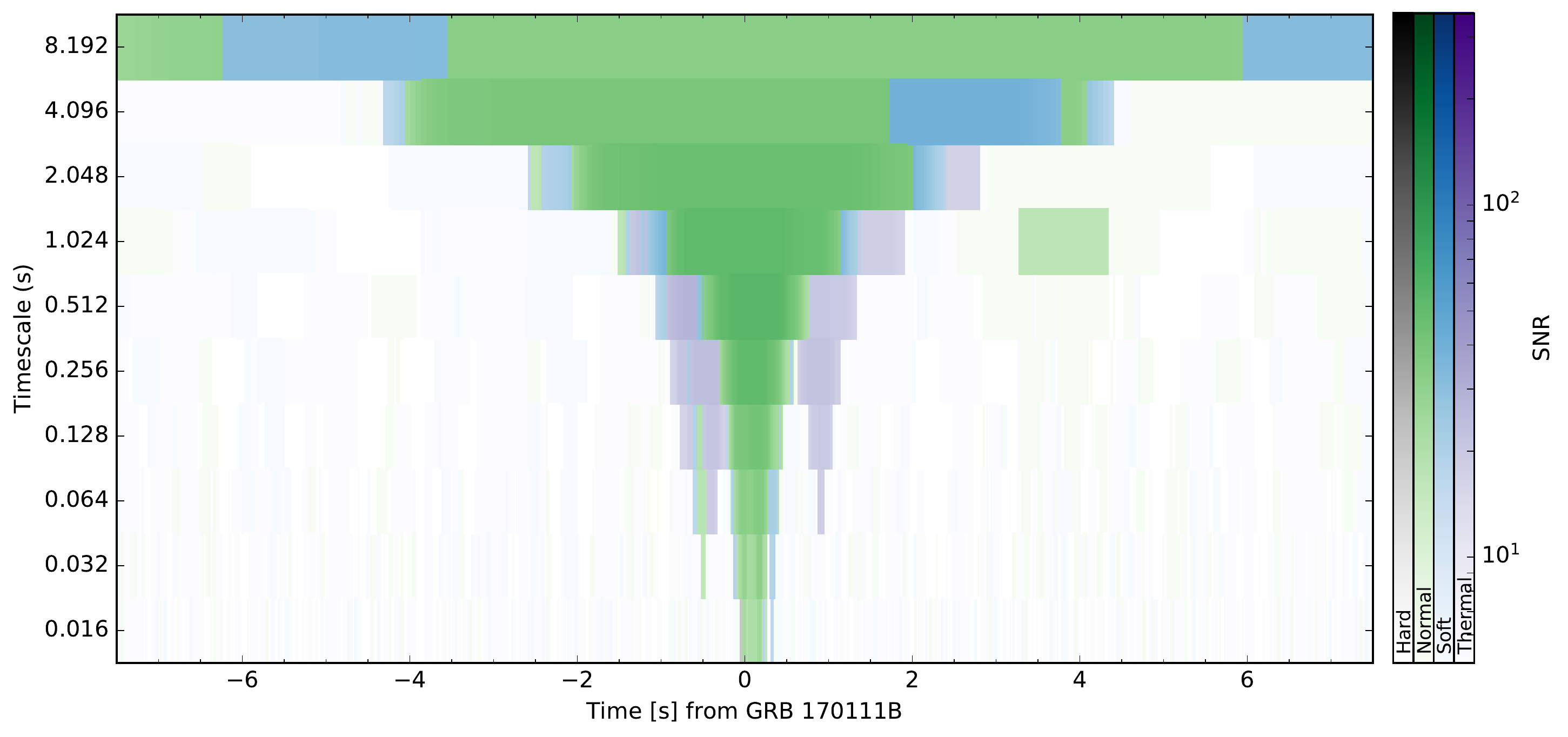} 
\label{fig:waterfall_plot}
\caption{Spectrally separated waterfall plots for GRB 130502A (top) and GRB 170111B (bottom) produced by the GBM Targeted Search. The GBM signal strength is colored by the best-fit spectral template, where the intensity is related to the signal-to-noise ratio of the bin.}
\end{figure}

The Targeted Search does not look for detector-coherent signals in counts space alone. 
The search also considers the deconvolved spectrum of the observed photons once folded through the GBM detector responses.
It assumes a set of four spectral templates each of which are folded through the 14 GBM detector responses, allowing for a search in deconvolved flux. The templates describing spectrally soft and medium GRBs are those mentioned in Section \ref{sec:localization}; however, the hard template is modeled as a PL with an exponential high-energy cutoff and parameters better suited for short GRBs \citep{Goldstein2016}. Motivated by the detection of the soft, thermal tail of GRB 170817A, a fourth BB template with a temperature of 10~keV was added (GBM Collaboration 2019, in prep). The BB template has not previously been used for signal detection, and thus we do not attempt to quantify the significance of the BB components with the Targeted Search. We apply it here for confirmation of spectral analysis and as follow-up characterization.

For each of these GRBs, the Targeted Search processed a 15 s window centered on the GBM trigger time, searching timescales ranging from 16 ms to 8.192 s by powers of 2. In each case, the main peak was found to be the event with the highest signal-to-noise ratio (S/N) within the search window. We identified the thermal component as the most significant BB bin found within the time range used in the spectral analysis. A summary of the search results, including start times, durations, and localizations for both  peaks and tails, is shown in Table \ref{tab:search_results}. The soft tails are weaker compared to the main peaks. However, the BB components of GRB 130502A and GRB 170111B were found to have S/N  of 8.54 and 7.77, respectively (see Table \ref{tab:search_results}), both of which are comparable to the main peaks of the sub-threshold short GRBs \citep{2018ApJ...862..152K}. 

The Targeted Search localizations of the main peak and soft tail for two candidate GRBs are shown in Figure \ref{fig:Targ_loc}. We quantify the spatial overlap between the two localizations as, 
\begin{equation}
{ \mathcal S } = \frac {{\sum}_{i=1}^{N}{P}_{1i}{P}_{2i}} {{\sum}_{i=1}^{N} {P}_{1i}{P}_{2i} + {\sum}_{i=1}^{N} {P}_{1i}u + {\sum}_{i=1}^{N} u{P}_{2i} } \, ,
\end{equation}
where $P_1$ ($P_2$) is the localization probability map for the main peak (soft tail), $i$ is the HEALPix \citep{gorski2005healpix} pixel index (corresponding to a sky location vector), and $u$ is the uniform probability for each pixel on the sky. Pixels of sky positions occulted by the Earth were excluded. Excluding GRB 140129A, the localizations for all soft tails identified with the Targeted Search were consistent with those of the main peaks, and this is reflected in the spatial probabilities listed in Table \ref{tab:search_results}. For GRB 140129A, the inconsistency in the localization of both the main peak and soft tail, as well as their poor spatial agreement, confirmed initial indications that GRB 140129A is not a likely candidate. Additionally, the emission preceding the main peak of 170111B was located to a centroid of (R.A., decl.) = (249.9, 54.7), verifying its spatial association with the burst.

\begin{figure}[b!]
    \centering
    \includegraphics[width=0.8\textwidth]{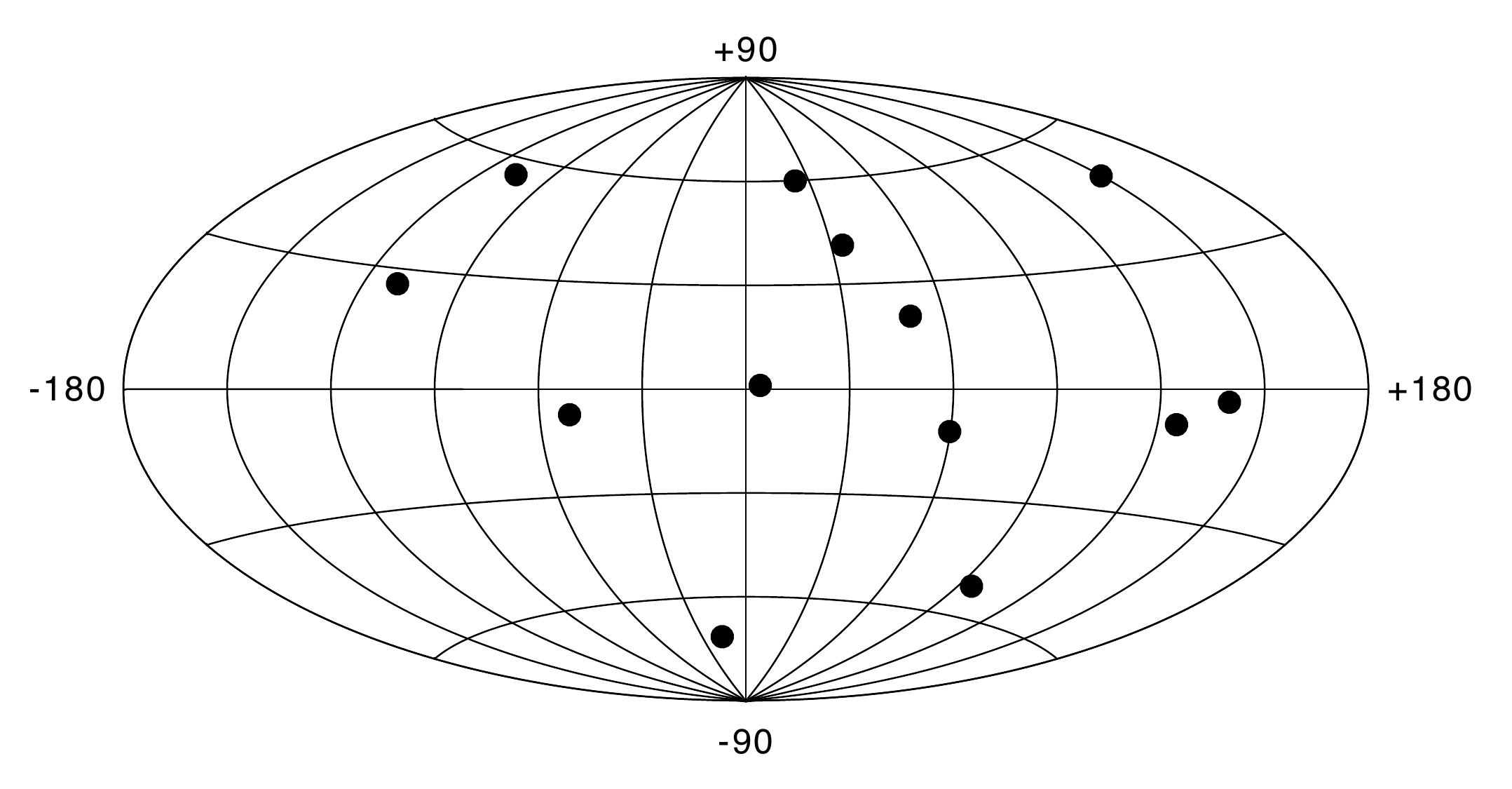}
    \caption{Locations of the 13 GRBs of Table \ref{tab:candlist_temp} in supergalactic coordinates.}
    \label{fig:SkyMap}
\end{figure}

Furthermore, the Targeted Search also produces spectrally separated waterfall plots that can act as proxies for detector light curves, with the addition of preliminary spectral information. For each timescale searched, the plot shows every data bin colored by the best-fit spectral template, where the color intensity is related to the S/N of that bin. We show these plots for GRBs with higher S/Ns, namely, GRB 130502A and GRB 17011B, in Figure \ref{fig:waterfall_plot}. In both GRBs, the thermal tails are clearly identified. While the thermal tail of GRB 130502A appears as a distinct component separate from the main emission, the thermal tail of GRB 170111B occurs closer in time to the main peak. GRB 170111B also shows the soft peak preceding the main emission.

\clearpage
\subsection{Anisotropy Tests}
The local galaxy density has a significant overdensity that appears as a plane on a two-dimensional map, the supergalactic plane. This structure is actually more complex, consisting of filaments, sheets, and knots, and is strongest at $\sim$40 Mpc and extends to $\sim$80 Mpc \citep{SGP}.    Since \grb is at 40 Mpc, it is possible that GRBs selected to be similar are mostly located close enough to show deviations from isotropy.  However, GRB~150101B having a redshift of 0.134 \citep{GCN17281} is significantly more distant.  Figure~\ref{fig:SkyMap} shows the locations of the final sample of 13 GRBs (Table~\ref{tab:candlist_temp}) in supergalactic coordinates -- the locations appear isotropic and there is no excess at the supergalactic plane. Quantitative tests of large-scale isotropy were made, checking the quadrupole moment about the supergalactic plane, $\langle \sin^2 {\rm SGB} - \frac{1}{3} \rangle$, and two coordinate-system independent statistics, the Rayleigh--Watson dipole statistic ${\cal W}$ and the Bingham quadrupole statistic ${\cal B}$ \citep{isobatse,isotheory}. All of these had values consistent with isotropy. Because the sample is very small, only a very strong anisotropy could have been detected.

\begin{figure}[h!]
    \centering
    \includegraphics[width=0.6\textwidth]{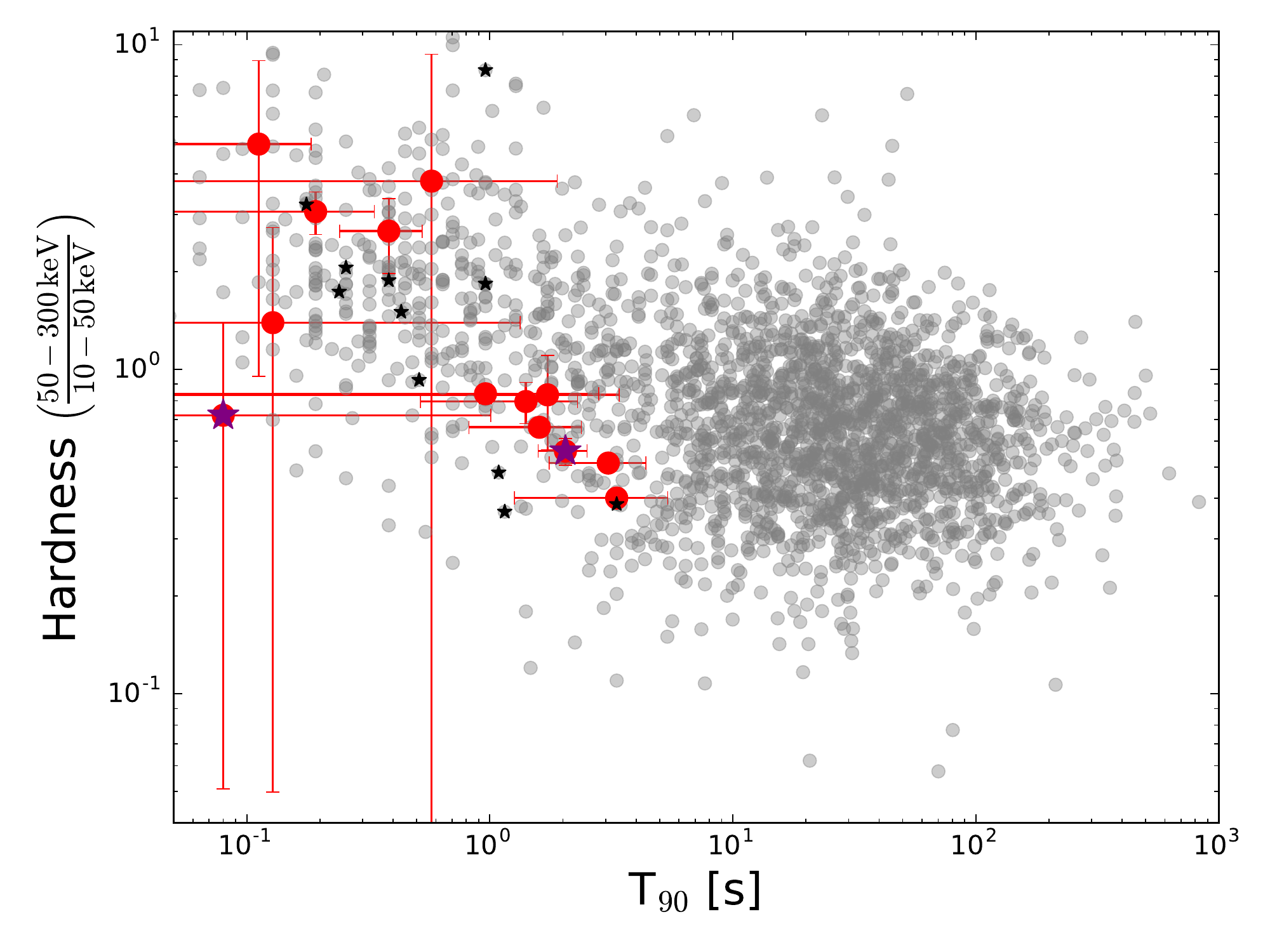}
    \caption{Hardness--duration distribution of \Fermi-GBM GRBs (gray). Red marks are the candidates, and stars indicate short ($< 5$~s) GRBs with measured redshift. Only GRBs 170817A and 150101B have measured redshift of the candidates (purple stars).  }
    \label{fig:HR-Duration}
\end{figure}

\subsection{Spectral Hardness vs. Duration}

Figure \ref{fig:HR-Duration} shows the hardness--duration plot from the 1ß yr GBM GRB catalog. The spectral hardness was obtained using the standard duration analysis performed in the catalog~\citep{GRB_Catalog_GBM_3rd_Bhat}. In this case, the hardness is defined as the ratio of the deconvolved counts in two energy bands (10-–50~keV and 50-–300~keV) obtained during the \tn~ time interval. We use 10 keV as the lower limit instead of 8 keV used in Section \ref{sec:BayesBlockAnal} to facilitate comparison with previous catalogs. The locations of the final sample are shown on this diagram as red marks. \grb~ and GRB 150101B are highlighted by the purple stars and GRBs (with \tn $<$ 5~s) with a redshift  by black stars, respectively. It emerges that the 13 GRBs in our sample are distributed in two small groups; one which is located at the soft tail below a hardness ratio value of one for a duration range between 1 and 4 s (7 GRBs including \grb). The second group consists of six GRBs (including GRB 150101B), each with a duration of less than 0.6 s. The latter
are distributed over a hardness range between 0.7 and 6 and exhibit larger errors due to the fact that the automatic hardness calculation barely registers the pulse of the GRB because it is relatively short. 
For better assessment, we inspect the peak energy of this group -  used as a proxy for the hardness ratio.
We find that the main pulse of the short group of candidates (see Table \ref{tab:candlist}) has systematically higher peak energies compared to the longer population. 
For the same reason we assign GRB 150101B, which shows a hardness below one, to the second group owed to its high peak energy. Hence, we rather consider the two groups by a slanted cut than by a single limit in hardness or \tn.
Looking more carefully at individual GRBs within the short group we find that the four sGRBs at the top of the diagram (hardness $>$ 2) are the cases where the catalog \tn~ does not include the soft tail emission. This can be checked by comparing the \tn~ time range, marked by two vertical dashed lines in Figure \ref{fig:lc4} and \ref{fig:lc5} with the time interval used for spectral analysis highlighted as red bins. For GRB 081209A, GRB 100328A, GRB 110717A, and GRB 120915A, we find no overlapping time interval leading to an increased hardness compared to the other GRBs where at least
some of the \tn~ range is overlapping the soft tail emission. 

\begin{figure*}[t!]
\centering
\includegraphics[width=.47\textwidth]{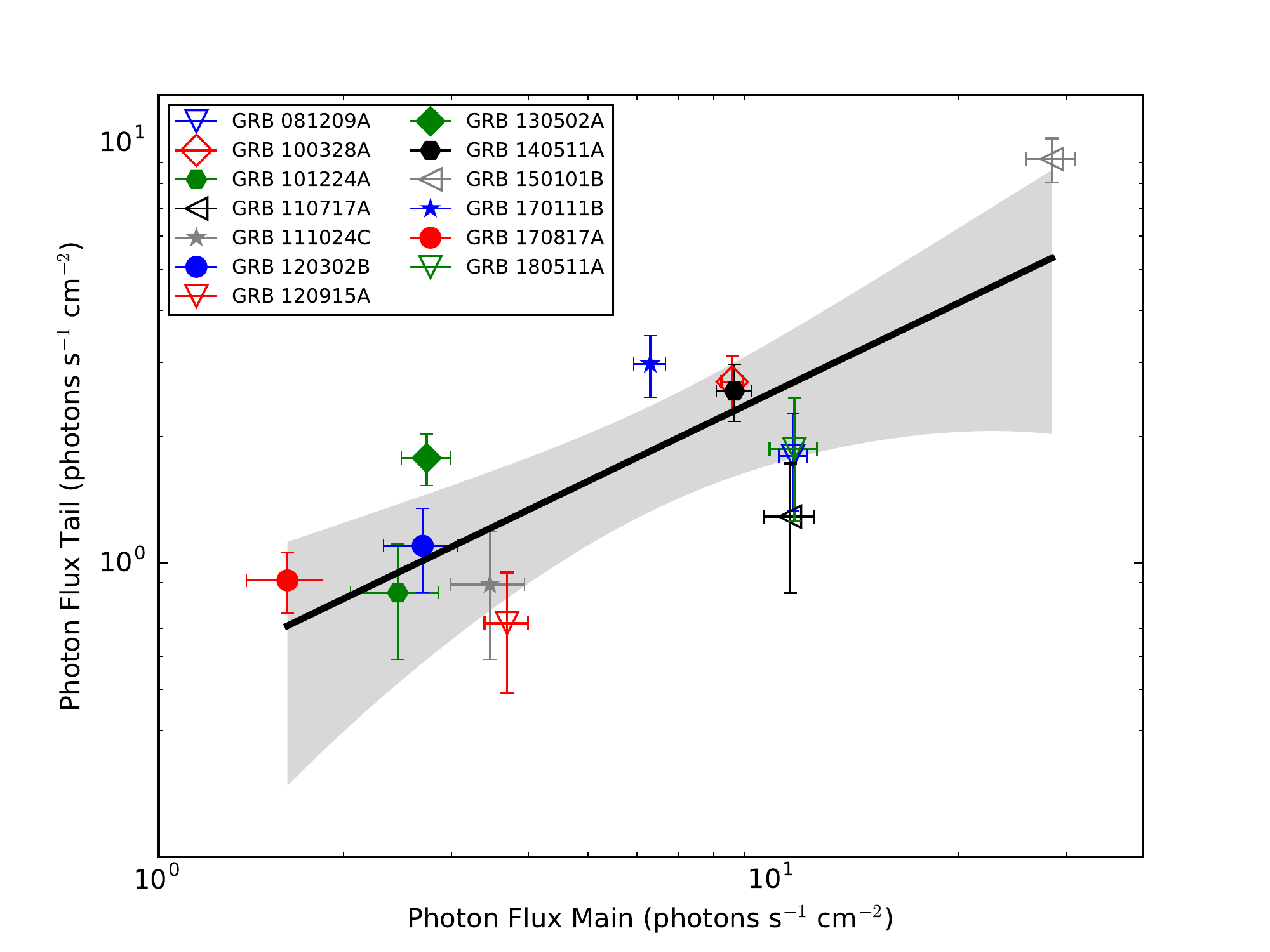}
\includegraphics[width=.47\textwidth]{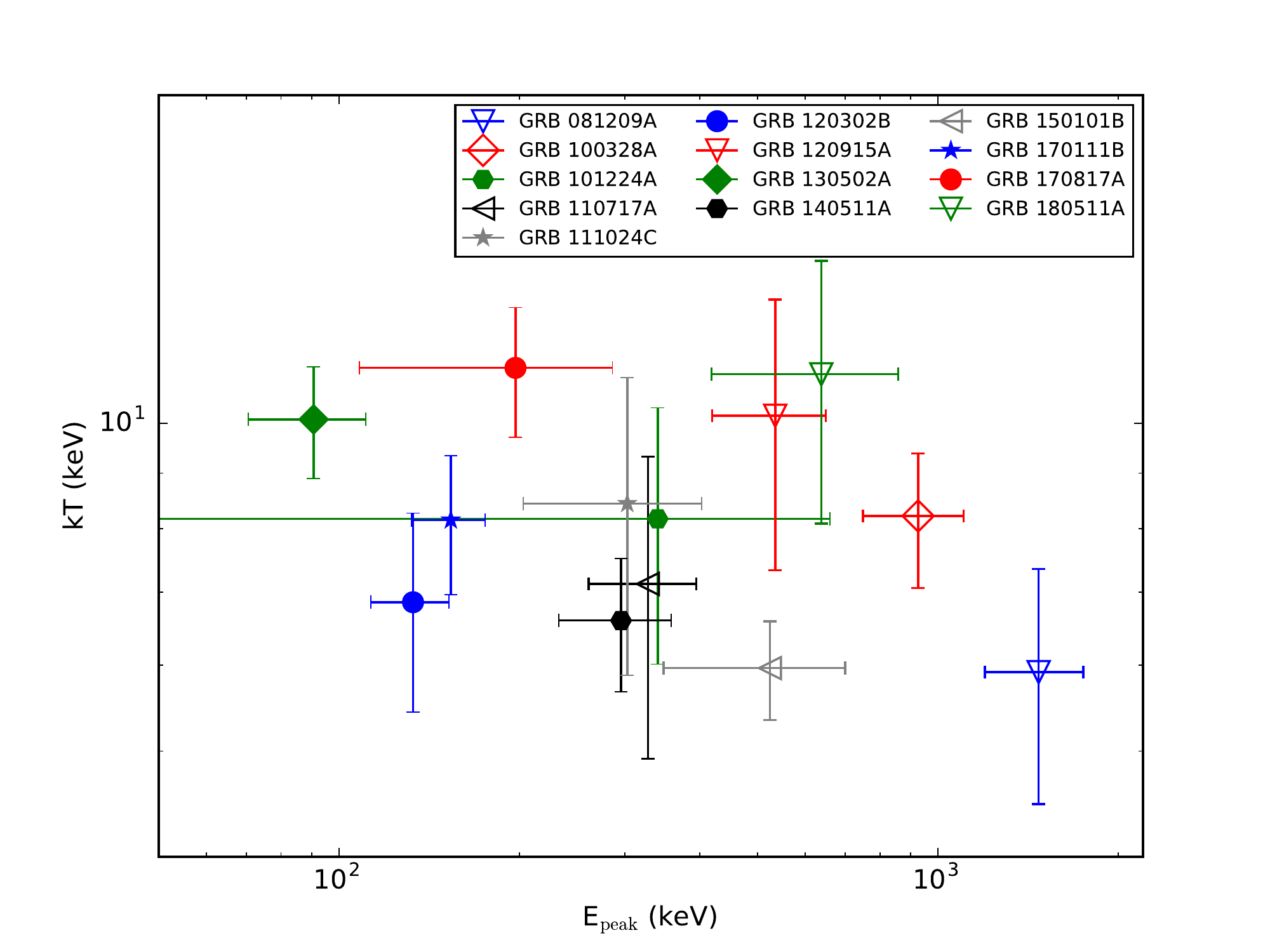}
\caption{Photon flux (left) and typical energy (right) for main peak versus soft tail. Hollow symbols mark the short-hard candidates, full symbols indicate  longer-softer  ones. The correlation between the photon fluxes is significant, the solid line is best power-law fit for the entire sample, and the shaded region is the 1 $\sigma$ confidence interval. There is no significant correlation between the characteristic energies.}
\label{fig:correl}
\end{figure*}

\begin{figure}[t!]
    \centering
\includegraphics[width=0.45\textwidth]{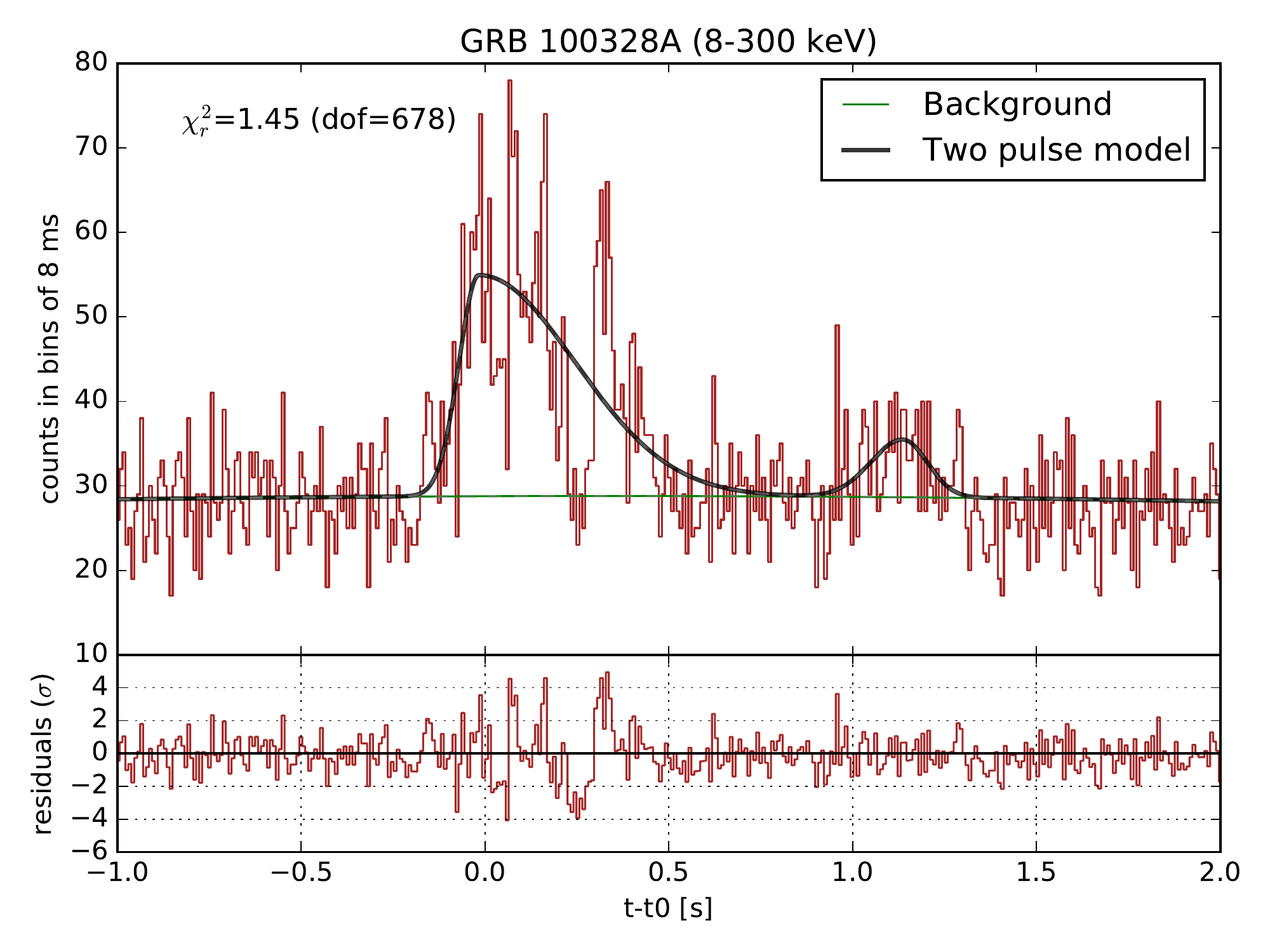}
\includegraphics[width=0.45\textwidth]{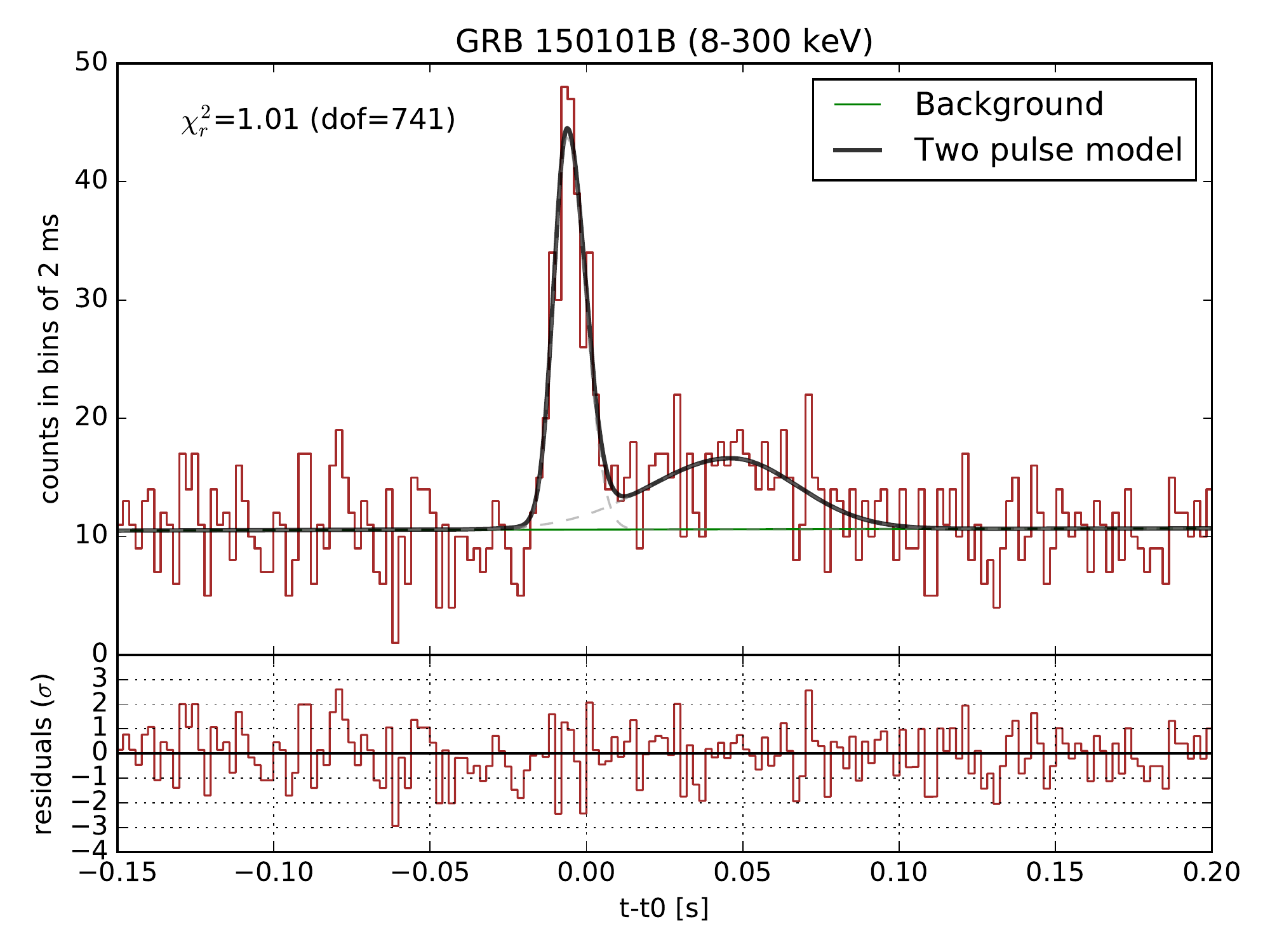}
\includegraphics[width=0.45\textwidth]{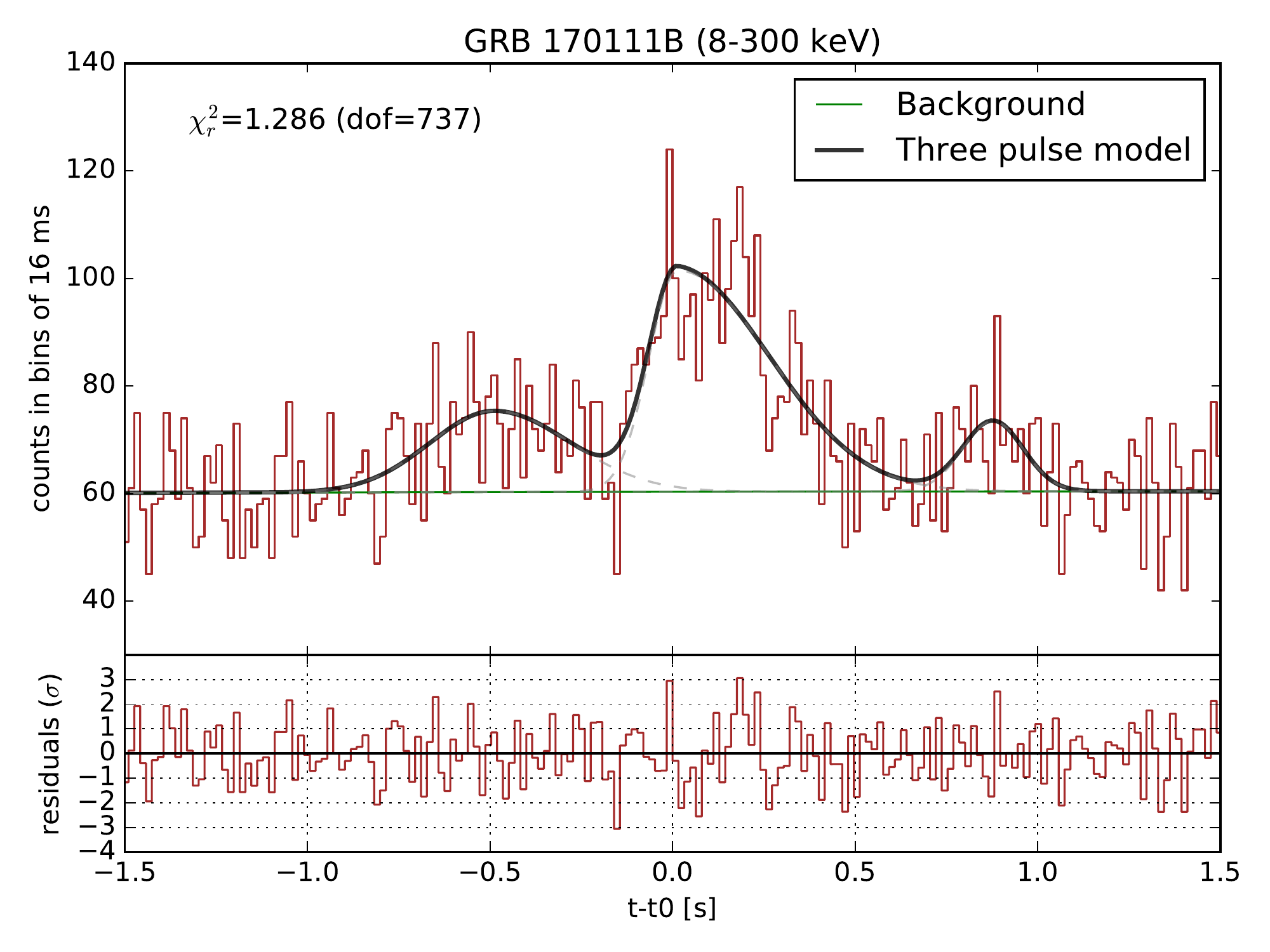}
\includegraphics[width=0.45\textwidth]{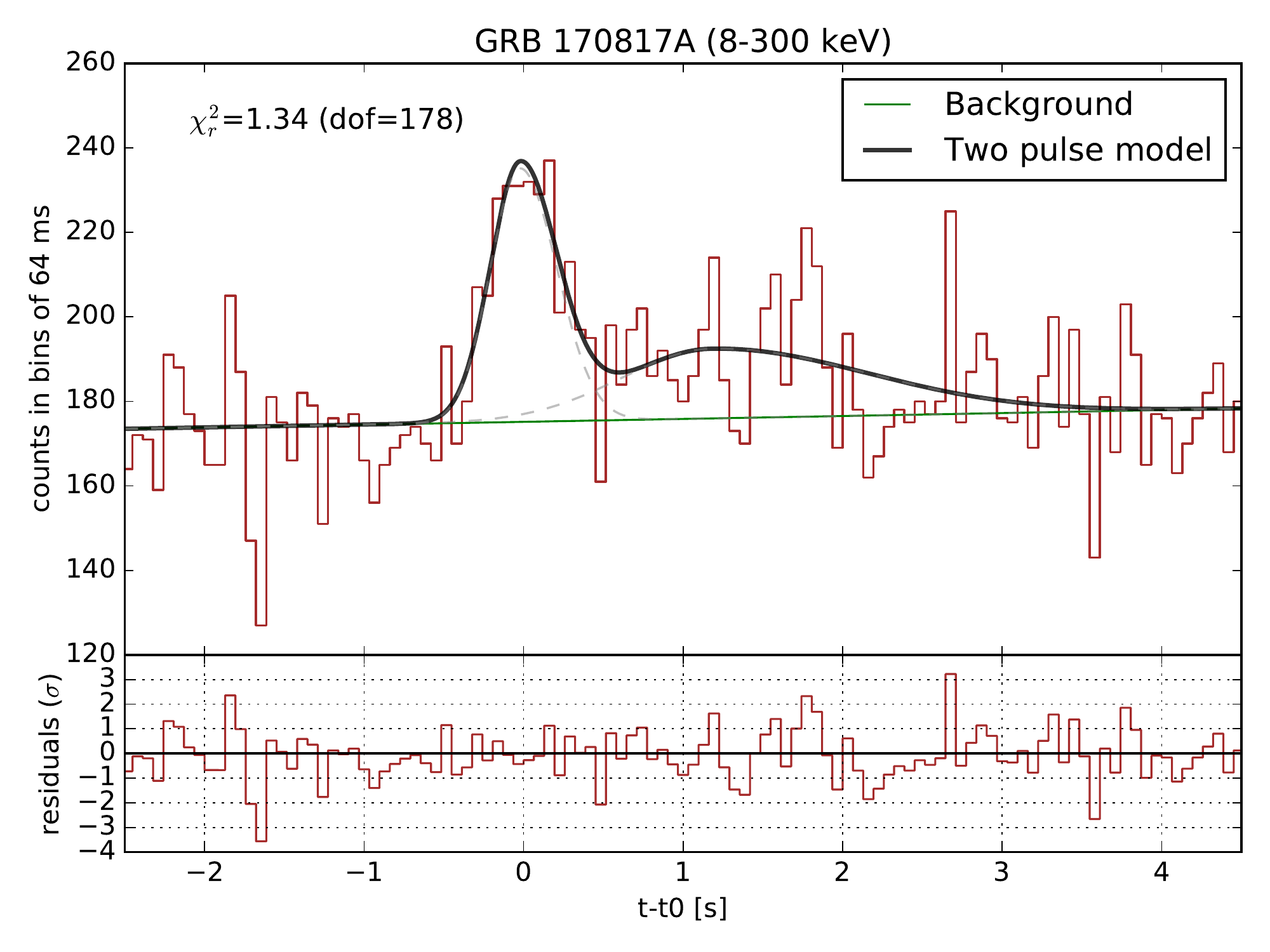}
    \caption{Examples of pulses in candidates to illustrate different morphologies. GRB 100328A shows multiple short pulses in the main pulse and it is separated from the soft peak (the one-pulse model fits the overall trend, but not the short timescale). GRBs 150101B and 170817A show the soft pulse starting approximately at the end of the main episode. GRB 170111B is peculiar in that it has soft emission both before and after the main pulse.}
    \label{fig:lcpulse}
\end{figure}

\subsection{Correlation Analysis}
We have explored correlations between parameters of the main pulse and soft tail. These included photon and energy fluxes, fluence and characteristic energy ($kT$ and $E_{\rm peak}$). We find a significant, Spearman rank correlation coefficient of $C=0.67$ in the photon fluxes. The $p$-value, or the probability that $C > 0.67$ correlation occurs by chance is 0.012. A PL fit to the photon flux gives ${P}_{\rm Main}\propto{P}_{\rm Tail}^{0.70\pm0.16}$ (see Figure \ref{fig:correl}, left). 

We find no significant correlation between the fluence and characteristic energies either for the whole sample or within the two subgroups. We plot the soft tail's temperature as a function of the main peak in the right panel of Figure \ref{fig:correl}.

The correlation in the photon flux points to the fact that the luminosities of the main episode and the soft tail may be linked rather than the total energetics.

\subsection{Pulse Fitting and Variability}

To further characterize the relation of the main pulse to the tail, we inspect the light curve using pulse-fitting techniques. We find cases where the two episodes are clearly separated, and some that overlap (see Figures \ref{fig:lcpulse}, \ref{fig:lc4}, \ref{fig:lc5}). 

GRB pulse shapes can be well described by analytical functions~\citep{Norris+96pulse,Norris+05pulse}. 
We fit a function composed of two pulses to the light curve in the 8--300~keV energy range, in order to study the properties of the main pulse and the soft emission.
In \citet{Norris+96pulse} a single pulse shape is given by $I(t)=A \exp{(-((t_{\rm peak}-t)/\sigma_{\rm rise})^2)}$ for $t<t_{\rm peak}$ and $I(t)=A \exp{(-((t-t_{\rm peak})/\sigma_{\rm decay})^2)}$ for $t>t_{\rm peak}$, where $A$ is the amplitude at the peak time of the pulse, $t_{\rm peak}$, $\sigma_{\rm rise}$ and $\sigma_{\rm decay}$ are the characteristic rise and decay times of the pulse respectively. 

The minimum variability timescale, $dt_{\rm min}$, describes the shortest coherent variation in the light curve, and can be used to infer the radius of the emission region. For the candidates, we determine $dt_{\rm min}$ based on the method of \citet{Min_Variability_Golkhou_2015}.  
As the main pulse carries most of the flux, the variability timescale informs us whether there are any significant structures shorter than the typical timescale of the pulse (e.g.\ the rise time). Based on the temporal parameters of the main pulse, we find that short-hard candidates with the exception of GRB~150101B have significant variation within the main pulse, i.e., they are composed of multiple overlapping pulses. 
We indicate the properties in this subsection in  Table \ref{tab:temporal}.

\begin{table}[t!]
\centering
\caption{Temporal parameters of the candidates all in units of $ms$. The (v) mark indicates candidates, where the variability timescale is less than the rise time with more than 2 sigma significance, indicating pulse sub-structure.}
\begin{tabular}{l|c|c|c|c|c}
\hline \hline 
GRB name         &  $\sigma_{t_{\rm rise,main}}$    & $dt_{\rm min}$ & $t_{\rm peak,soft}$ - $t_{\rm peak,main}$ & Main/Tail Relation\\
\hline         
GRB 081209A (v)&    25 $\pm$ 3       & $ <$ 14.9            & 133 $\pm$ 14   & joined \\ 
GRB 100328A (v)&    77 $\pm$ 21        & $ <$ 10.6         & 1153 $\pm$ 65  & separated \\
GRB 101224A &    23 $\pm$ 23           &  47.4 $\pm$ 7.6   & 1360 $\pm$ 625 & separated \\
GRB 110717A (v)&    36 $\pm$ 11       &  11.4 $\pm$ 3.0    & 712 $\pm$ 2097 & separated \\
GRB 111024C &    33 $\pm$ 17          &  40.7 $\pm$ 8.8    & 106 $\pm$ 20   & separated \\
GRB 120302B &    16 $\pm$ 19           & $ <$ 119.6        & 1545 $\pm$ 134 & separated \\
GRB 120915A (v)&    312 $\pm$ 75       &  40.6 $\pm$ 13.2 & 632 $\pm$ 717  & separated \\
GRB 130502A &    169 $\pm$ 63          &  220.7 $\pm$ 34.0& 2092 $\pm$ 765 & separated \\
GRB 140511A &    23 $\pm$ 7            & $ <$94.4           & 385 $\pm$ 424  & joined \\
GRB 150101B &    6 $\pm$ 1           &  7.9 $\pm$ 0.7        & 52 $\pm$ 11    & joined \\
 &     &           & 865 $\pm$ 71 &  separated\\
\rb{GRB 170111B} &    \rb{110 $\pm$ 36}     & \rb{$ <$63.4}         & -502 $\pm$ 104 \tablenotemark{a}  & separated    \\
GRB 170817A &    263 $\pm$ 103  &   124.6 $\pm$ 6.4 & 1201 $\pm$ 774 & joined \\
GRB 180511A (v)&    15 $\pm$ 4       & $ <$5.3           & 94 $\pm$ 65   & joined  \\
\hline
\end{tabular}
\label{tab:temporal}
\tablenotetext{a}{Result for the pretrigger soft emission of GRB 170111B }
\end{table}
%

\subsection{Comparison with GRB 170817A}
\begin{figure}[ht!]
    \centering
\includegraphics[width=0.42\textwidth]{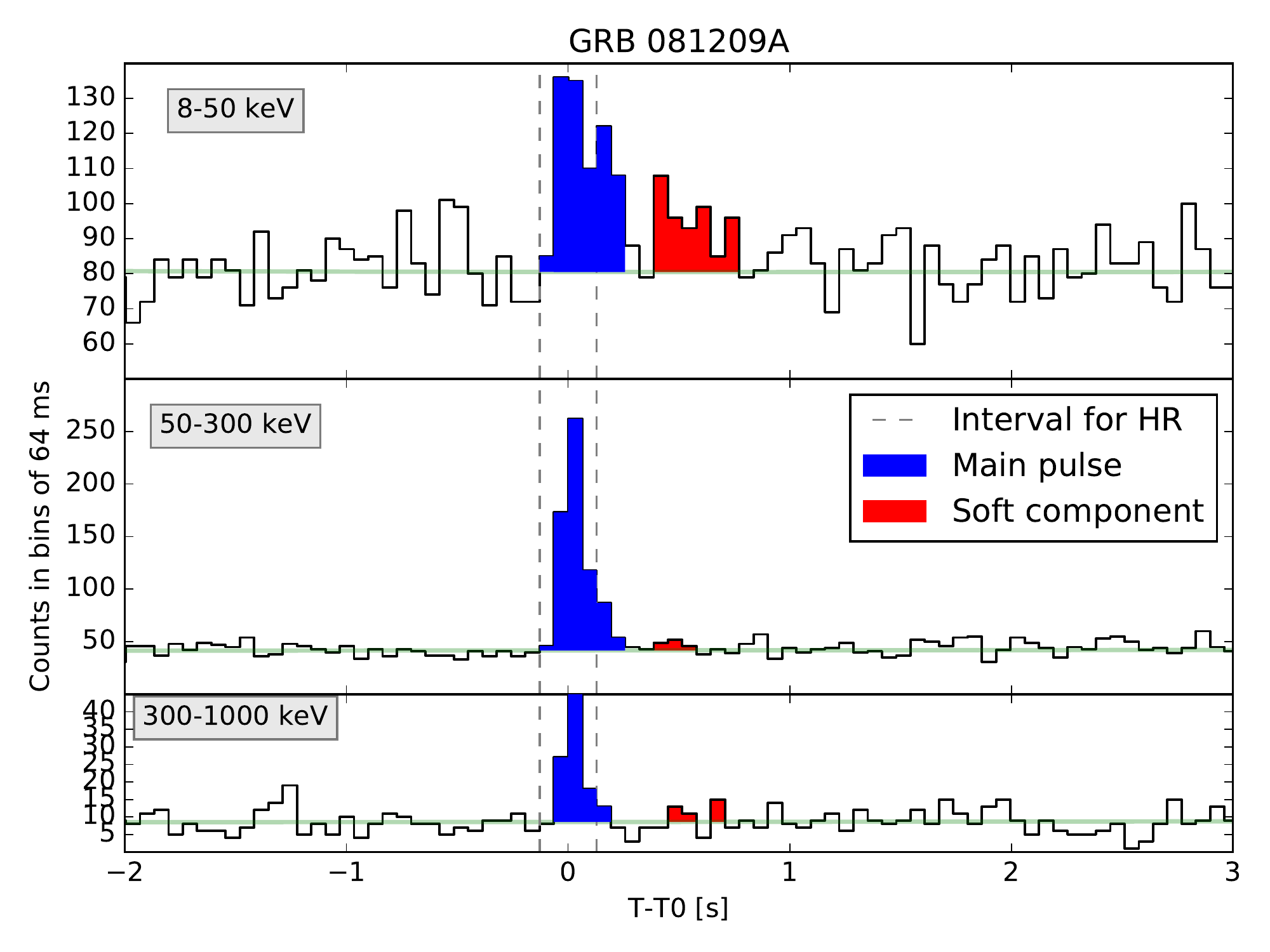}
\includegraphics[width=0.42\textwidth]{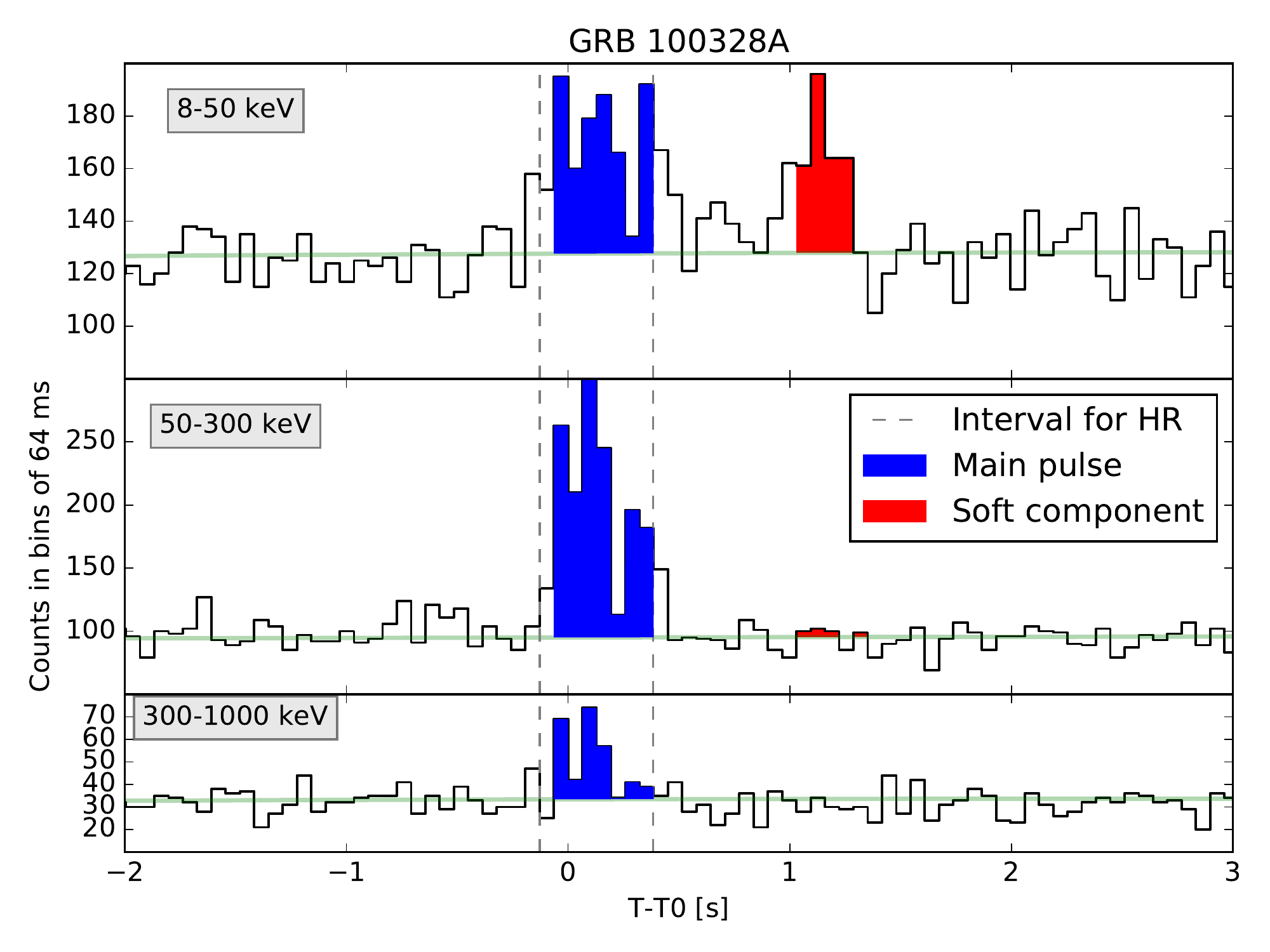}\\
\includegraphics[width=0.42\textwidth]{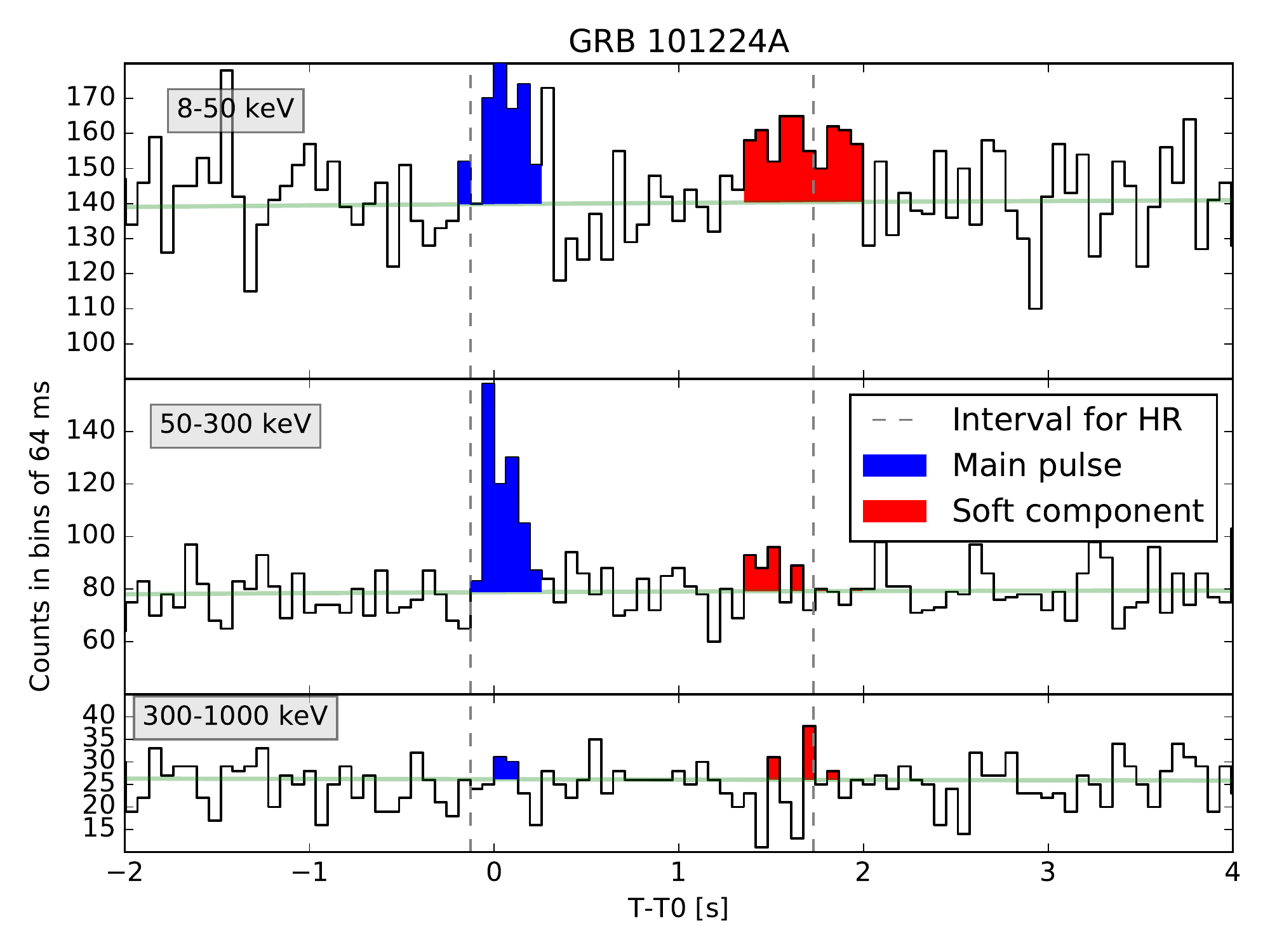}
\includegraphics[width=0.42\textwidth]{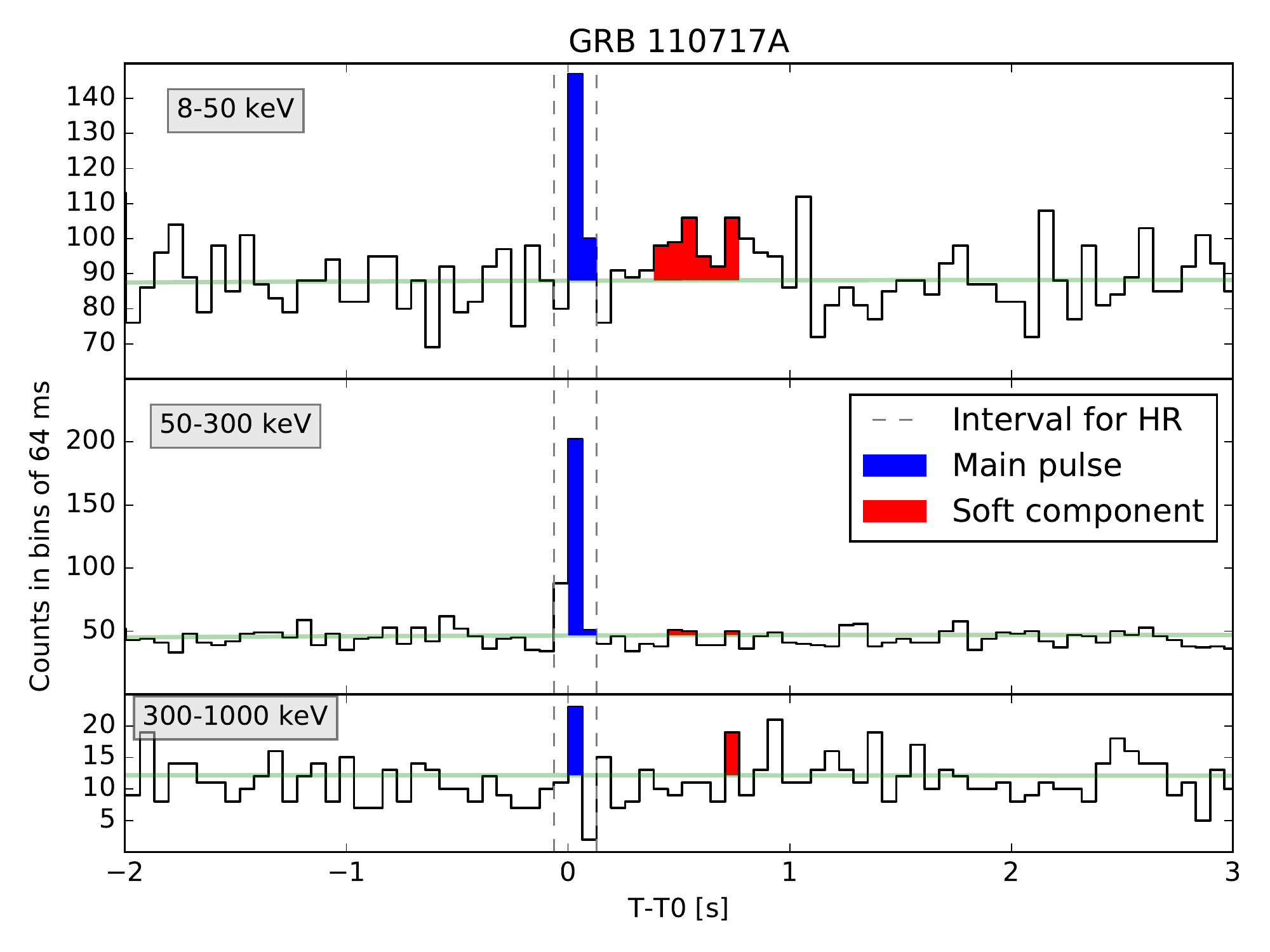}\\
\includegraphics[width=0.42\textwidth]{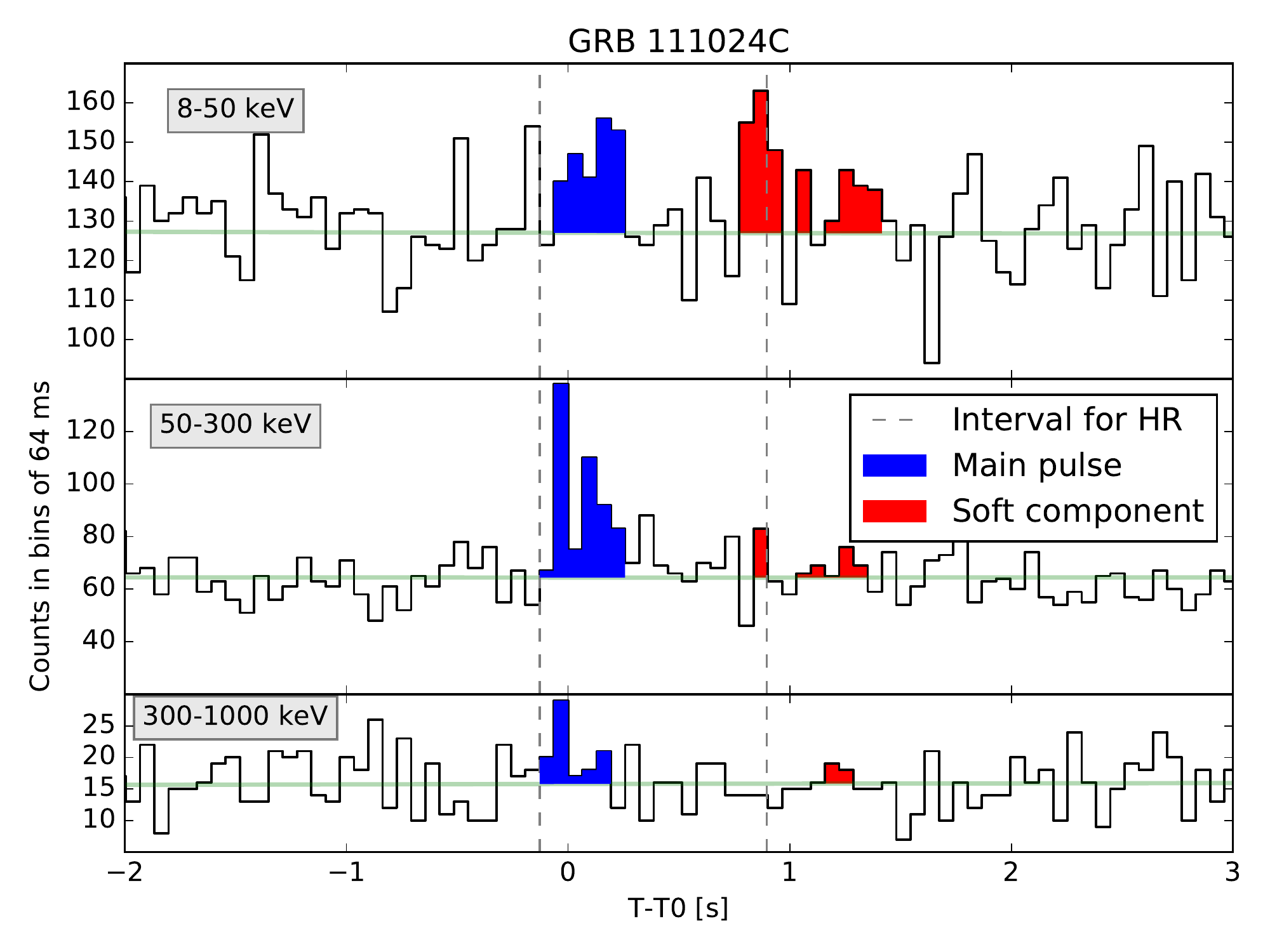}
\includegraphics[width=0.42\textwidth]{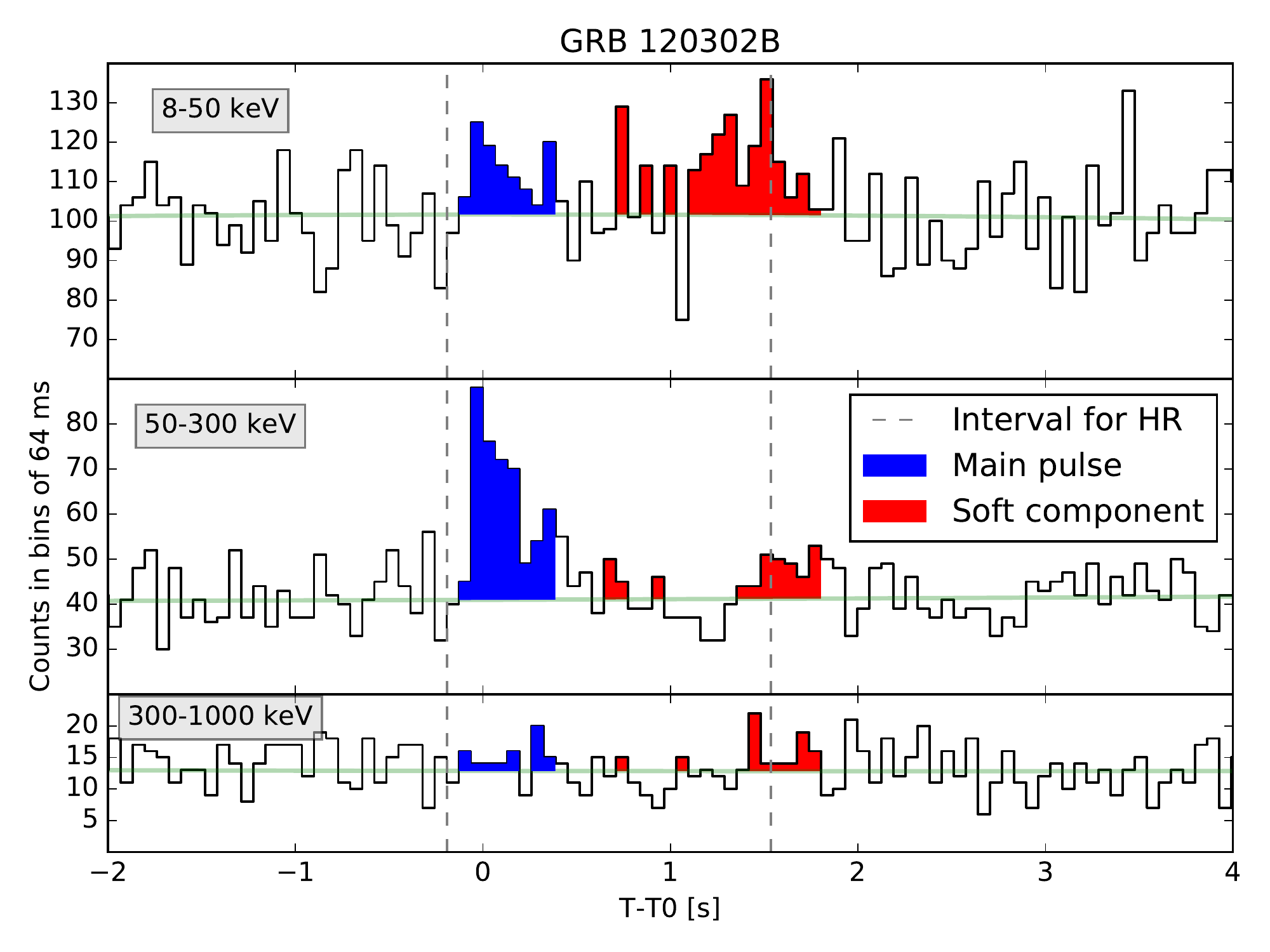}\\
\includegraphics[width=0.42\textwidth]{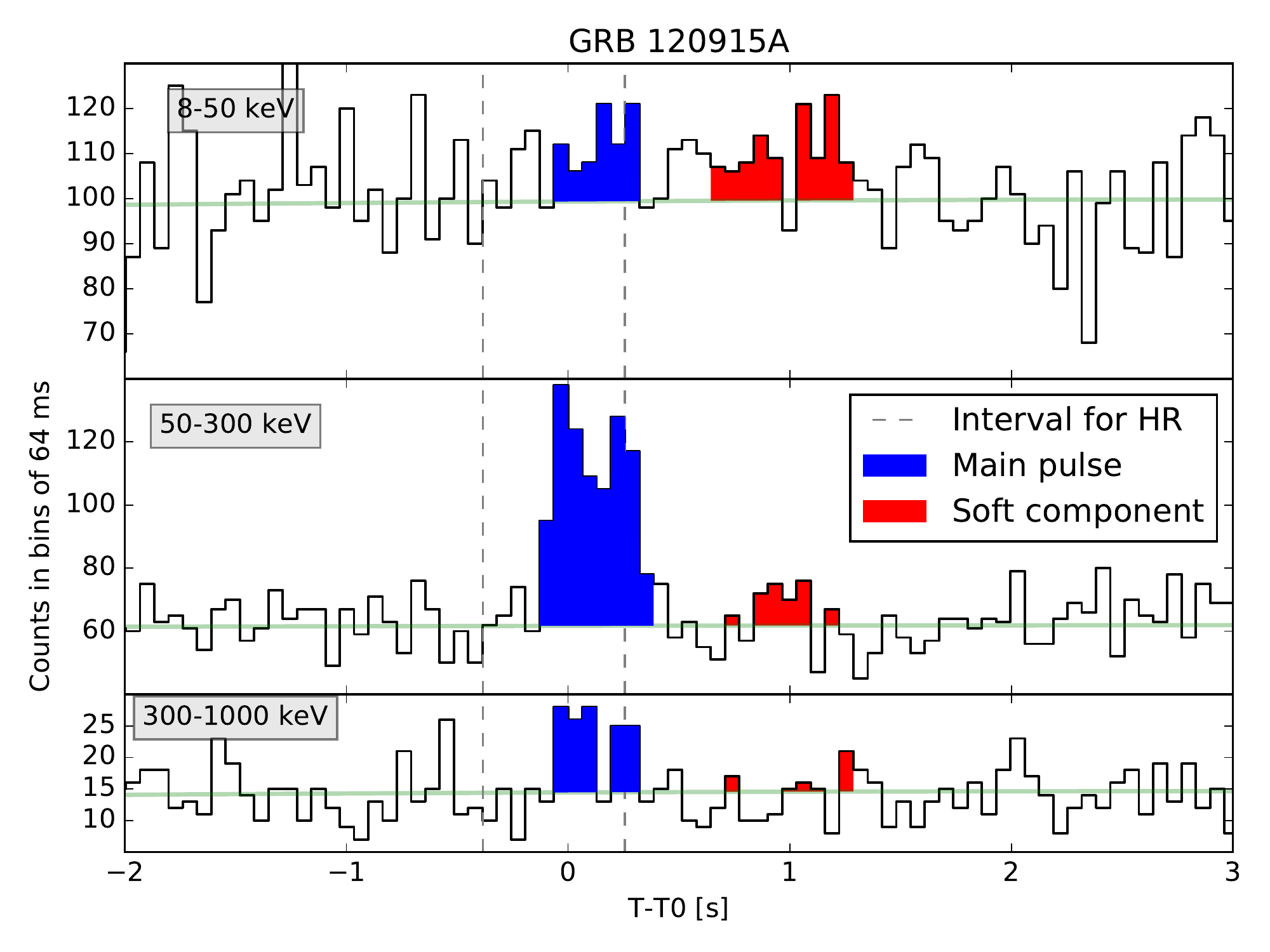}
\includegraphics[width=0.42\textwidth]{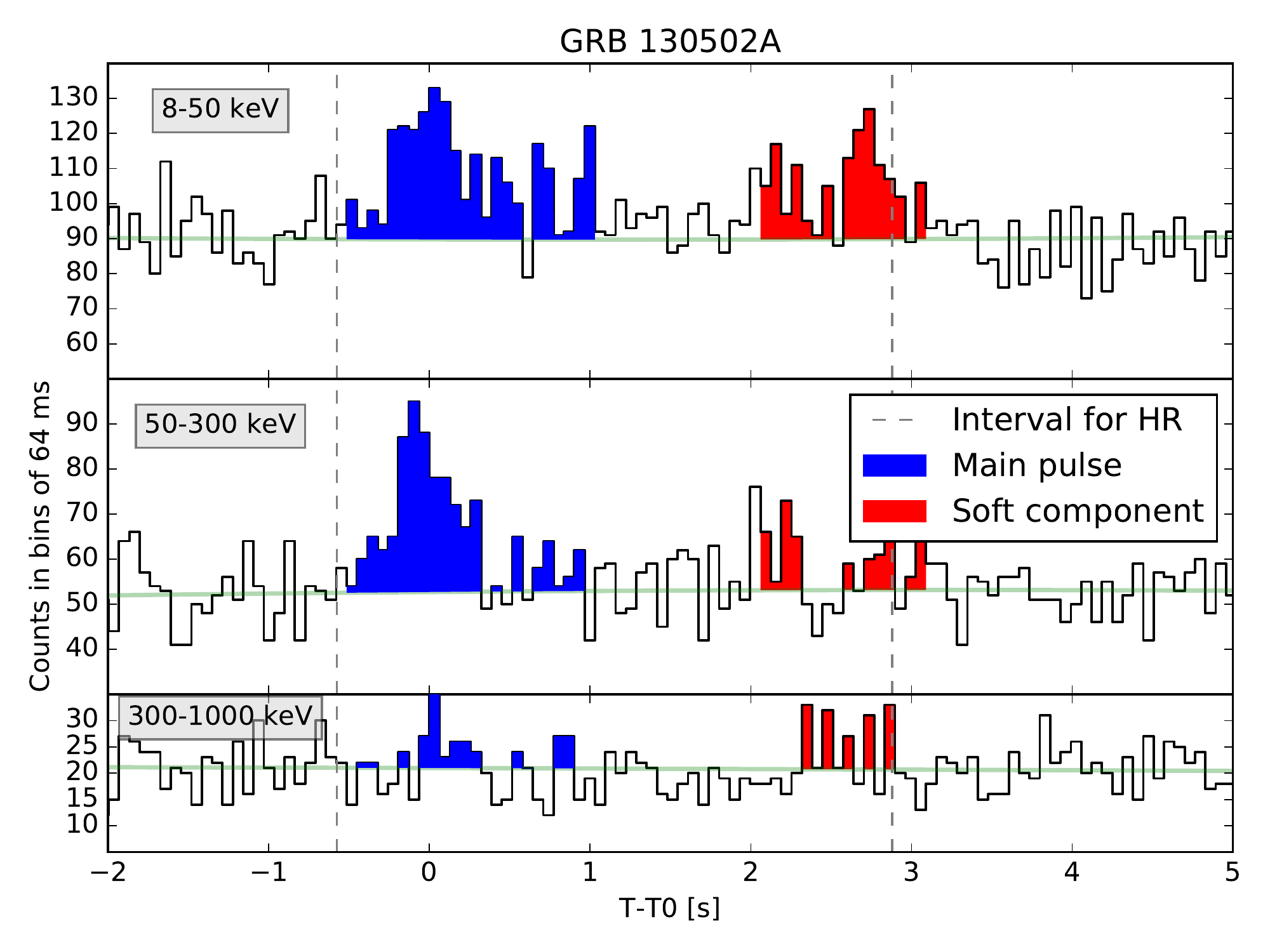}
    \caption{Three channel light curves of the candidates. We indicate the main pulse used for the spectral analysis with blue and the soft emission with red.  These intervals might not exactly correspond to the ones from the Bayesian analysis}. The interval, determined automatically to calculate the hardness ratio, is marked with dashed vertical lines. Green line marks the background.
    \label{fig:lc4}
\end{figure}

\begin{figure}[ht!]
    \centering
    \includegraphics[width=0.45\textwidth]{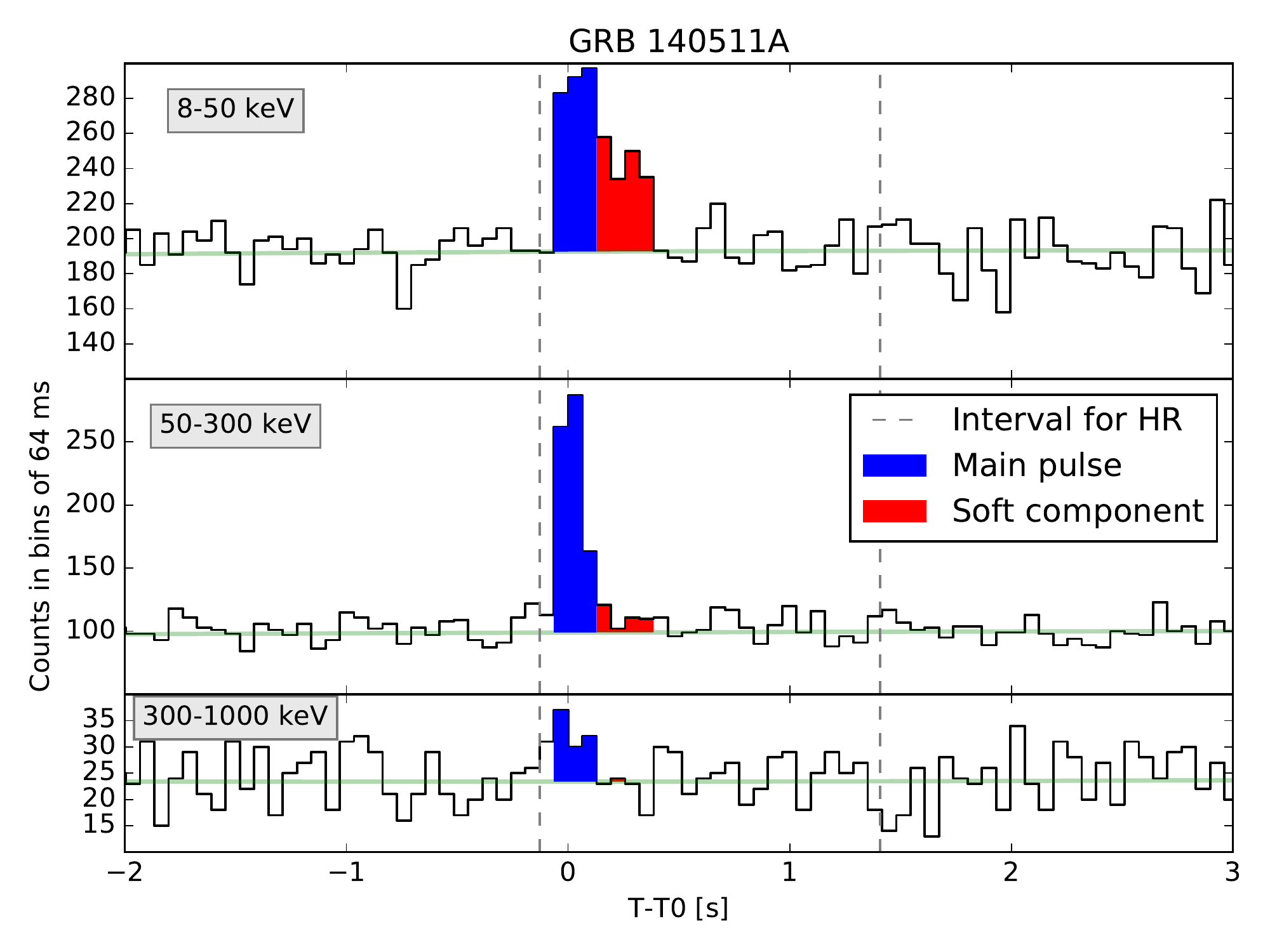}
\includegraphics[width=0.45\textwidth]{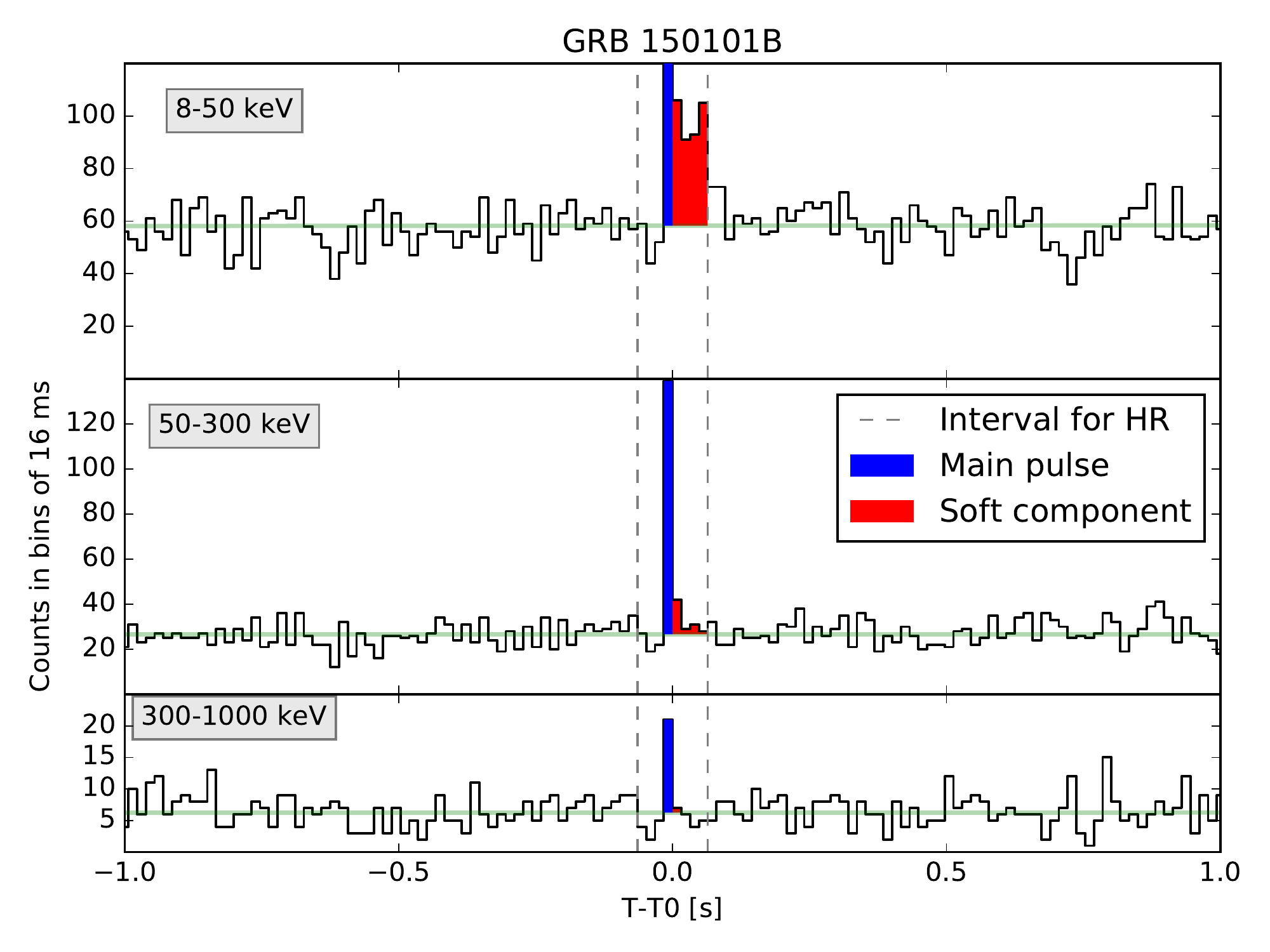}\\
\includegraphics[width=0.45\textwidth]{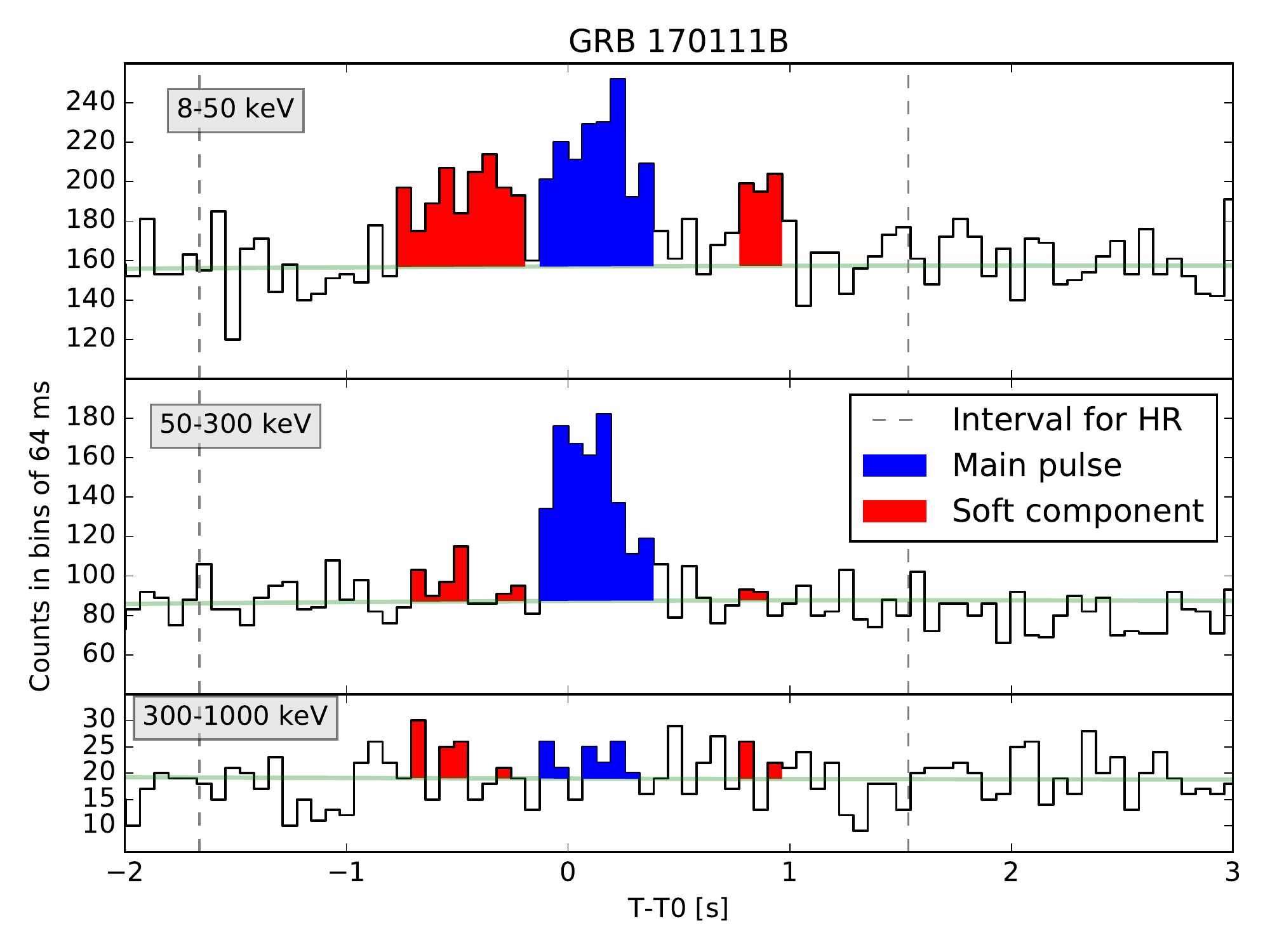}
\includegraphics[width=0.45\textwidth]{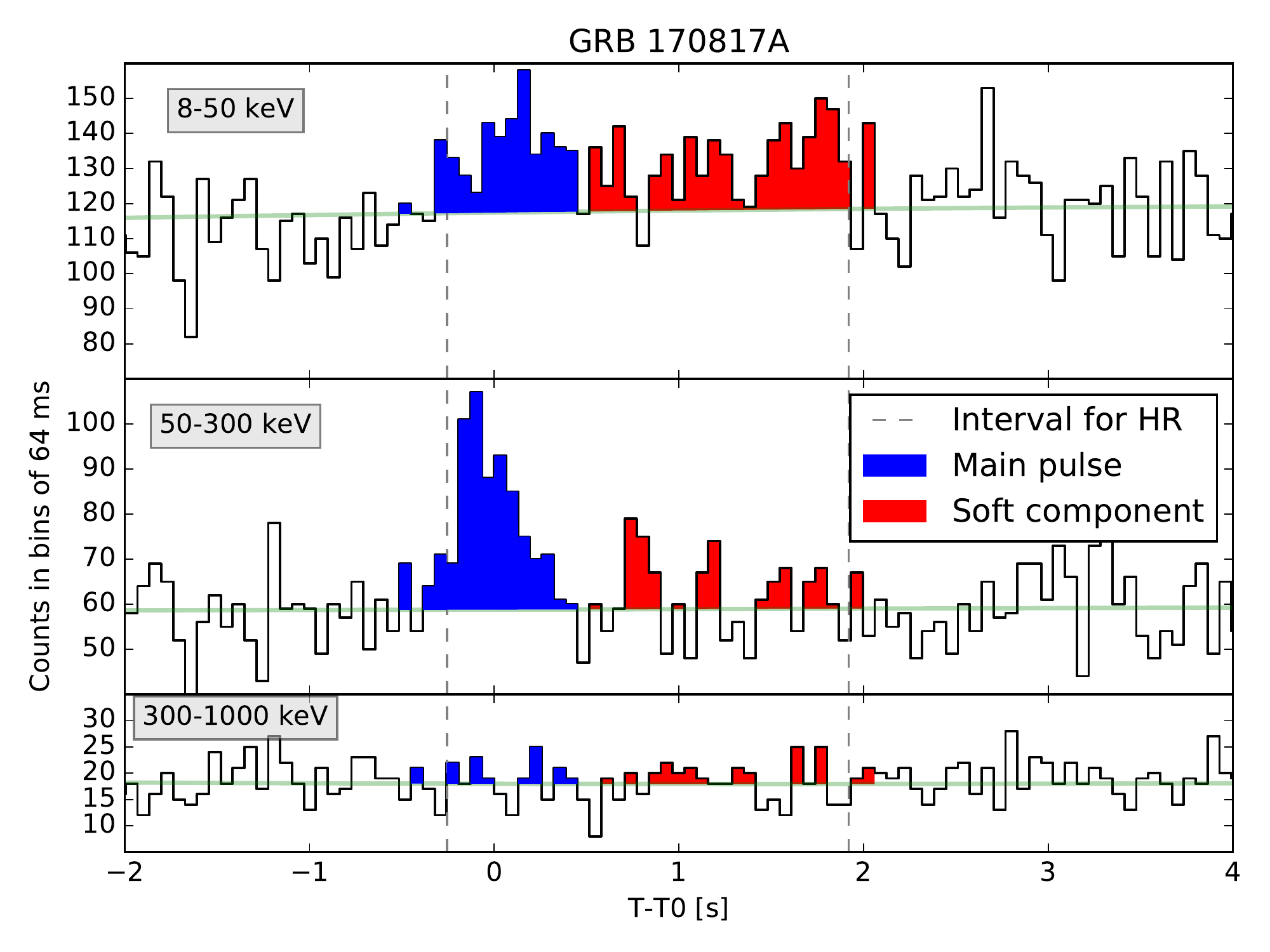}\\
\includegraphics[width=0.45\textwidth]{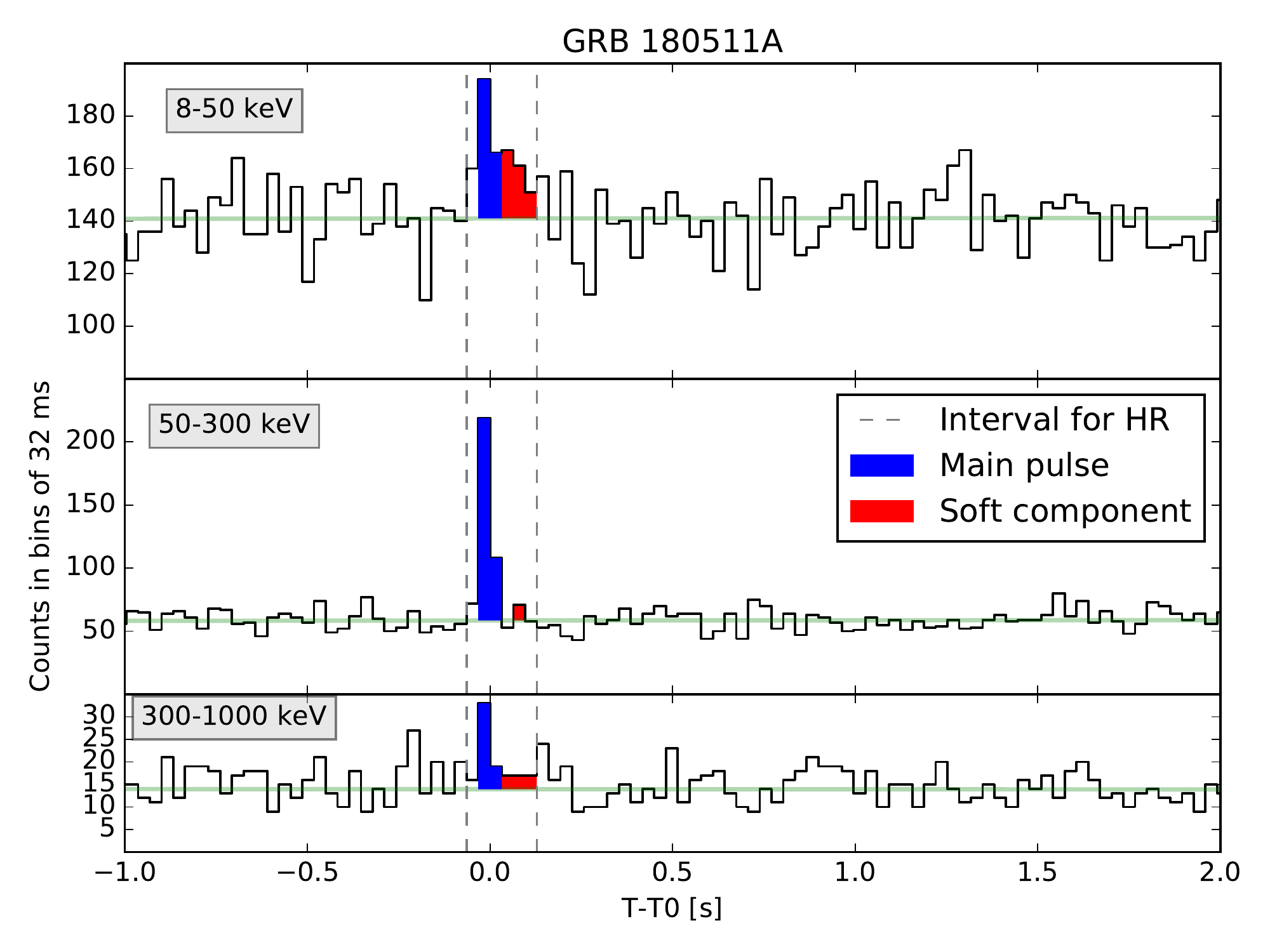}
    \caption{The light curves of the remaining candidates. Notations similar to Figure \ref{fig:lc4}.}
    \label{fig:lc5}
\end{figure}

All of these bursts can be described within the search criteria of the sample. They are all GRBs with a duration of less then five seconds, most of which can be described as a spectrally hard burst followed by thermal X-ray emission. The exception is GRB 170111B, which is the only burst to exhibit soft, thermal emission about 0.5~s before the main burst. The soft emission episodes for the final list of GRBs are seen prominently in Figures~\ref{fig:lc4} and~\ref{fig:lc5} and have BB temperatures of around 5--10~keV. The E$_{\rm peak}$ values for the main burst episode varies from around 100 to 1000~keV.

Only two bursts in our sample have recorded redshifts; GRB 150101B and GRB 170817A. To-date, these two bursts are among the closest known short-GRBs with redshifts to have been detected by \Fermi-GBM. These redshifts are 0.134 and 0.010 for 150101B and 170817A, respectively. The temporal arrangement between the main episode of the short GRB and the thermal emission is mostly what defines the morphology and duration of each burst. For a handful of bursts in the final sample, the soft component is found to be a temporally separate event from the main episode at high significance (GRB~100328A, GRB~101224A, GRB~120302B, and GRB~170111B). For the rest of the bursts there is no clear separation between the main pulse and the soft tail. We have to emphasize that we did not see any stark difference in our measured parameters between bursts with resolved soft components, and those that have no clear temporal separation.

\section{Discussion}
\label{sec:Disc}
We have selected a sample of GRBs that show similarities to \grb. We based our search on the presence of a soft emission episode with a BB spectrum that follows the main peak. BB spectra are commonly observed in bright GRBs 
(\citealp{ryde05}; \citealp[including sGRBs,][]{Guiriec+13shortthermal}); however, this component is always coeval with the main nonthermal phase. It was in the case of \grb, that soft emission separate from the main peak was reported for the first time. Based solely on the main pulse and soft tail structure, we have identified 13 GRBs in 10 yr of GBM observations. This translates to 1.3 similar GRBs per year. Only two of the candidates in our sample have measured redshifts, thus we cannot infer the luminosity or the energetics of the whole sample.

The two emerging groups of candidates based on the separation observed in the hardness--duration figure (see Figure \ref{fig:HR-Duration}) can be tentatively  explained as a viewing angle effect. Due to the Doppler effect, similar GRBs viewed off-axis will become softer and of longer duration. Moreover, the short timescale structures present in on-axis light curves will be smoothed out for an off-axis observer \citep{Salafia+16offaxis}. This is consistent with the location of the candidates on the hardness--duration diagram: short-hard candidates are viewed closer and more on-axis, while the softer and longer candidates are viewed more off-axis. The fact that the shorter GRBs show significant short timescale variability within their pulse also reinforces this picture (Table \ref{tab:temporal}).

There are two models that address the pulse and tail structure observed in the prompt emission. Testing these models requires detailed simulations, beyond the scope of this work. Here, we outline the main results from our study that models need to address. 
The merger of two neutron stars is accompanied by significant mass ejection either through tidal mechanisms or through winds, e.g.\ driven by the disk formed around the central remnant right after the merger. The relativistic jets that power GRBs need to burrow through this ejecta. 

In the cocoon shock breakout model \citep{cocoon_gottlieb} the GRB jet (that is either successful in breaking out of the ejecta or choked by it) powers a cocoon that produces a mostly isotropic emission when a shock breaks out of it. The appeal of this model is that it naturally produces two emission components during the shock breakout: a fast evolving, bright emission called the {\it planar} phase right after the breakout, and lasting until the emitting radius doubles. This is followed by a less luminous emission that evolves slower, called the {\it spherical} phase. This model explains inherently weaker GRBs, as in the case for GRB 170817A. For brighter GRBs, like GRB 150101B, the simple closure relations between total energy, duration, and typical photon energy are not satisfied \citep{2018ApJ...863L..34B}.
The strong variability observed for many candidates could not come from the shock breakout emission. One could possibly argue that a successful (on-axis) jet is providing the highly variable flux in the confines of this model, outshining the less luminous planar phase of the cocoon shock breakout and the soft emission is supplied by the spherical phase of the breakout emission. In the cocoon shock breakout model the two types of emission are produced by the same expanding shock, it is unclear how to account for the soft tails that are temporally clearly separated from the main pulse.

A different model \citep[e.g.][]{cocoon_lazzati_2016} ascribes the main peak to a successful GRB jet, with lateral angular structure \citep{Kathi+18offaxis} that is viewed off-axis. The soft emission comes from the photosphere of a wide-angled cocoon that develops similarly to the previous model. As it invokes the well-studied GRB emission mechanisms for the prompt emission (e.g.\ internal shocks), this model can explain both the highly variable main emission and the soft tail. This is done, however, at the price of introducing the cocoon photosphere as an additional component. 

GRB 170111B has an intriguing soft precursor that is difficult to understand in any of the simple variants of the above two models. A possible scenario would ascribe the first soft spectral component to thermal emission from a jet photosphere.  The nonthermal emission of this jet has to be suppressed (e.g.\ because of low efficiency), so the thermal component dominates. With additional activity from the central engine, the nonthermal main peak emerges, e.g.\ from interaction between additional jet components. Finally, another jet emerges with  suppressed nonthermal emission or with a cocoon to produce the second soft tail. Such a picture can be further refined or rejected by additional observations.

Based on our candidates any model explaining the main peak+soft tail structure needs to account for a correlation between the peak flux of the main pulse and the soft tail (and no correlation between the fluences).  This correlation indicates that the main link between the two components is the luminosity and not the total energy.


\section{Conclusion}
\label{Sec_Conc}

The detection of \grb in coincidence with GW170817, marked the start of joint GW-GRB detections. Based on its unique gamma-ray properties, we have identified 13 GRBs in the \Fermi-GBM catalog that are similar to \grb. 
In our selection, we employed different criteria for finding similar GRBs to \grb. The initial approach involved finding bursts within a range of measured parameters (duration, peak flux, etc.). We refined this approach by employing a new search based on a Bayesian Block method that identified emission excesses in the softer GBM detector channels. We finally analyzed the candidates manually. We vetted our list of candidates to ensure the locations of the main and soft emission episodes were similar, using the targeted search where possible. Consequently, we find a total of 13 candidates. The observation time span of 10 yr allows us to estimate the rate of these events at $\sim$ 1.3 per year detectable by GBM.

We further find our final of 13 candidates separate clearly into two groups on the duration--hardness plane. The shorter GRBs show fast variability, shorter than the size of the main pulse. These observations challenge current models.

One of the most interesting results is the prior soft emission found in GRB 170111B, where in addition to the soft tail, we found a soft precursor with a comparable flux and temperature. If the proposed candidates are indeed BNS counterparts, theoretical models will be required to account for an early soft emission as well as the late soft tail.

Having a confirmed short gamma-ray burst counterpart to a BNS merger puts the BNS scenario as the origin of sGRBs on solid footing, although other scenarios (BH--NS mergers or magnetars) could still viably account for at least a fraction of sGRBs.
Here, we have used the most striking feature of \grb (a nonthermal pulse followed by a thermal tail), to search for other similar events. In case this feature is an indicator of BNS origin for sGRBs, it is not clear if it is a universal gamma-ray property. This is because the soft tail is always weaker than the main pulse and therefore, a main pulse without a thermal tail may be observed for sGRBs occurring at larger distances. It is also possible that we have identified some special circumstance (e.g.\ viewing angle effect, proximity, binary component masses, etc.)  that produces the ``pulse + tail" structure and the bulk of the BNS mergers produce typical sGRBs (without the soft tail). We expect future joint GRB/GW observations will clarify this picture.

\section*{Acknowledgements}

Support for the German contribution to GBM was provided by the Bundesministerium f{\"u}r Bildung und Forschung (BMBF) via the Deutsches Zentrum f{\"u}r Luft und Raumfahrt (DLR) under grant number 50 QV 0301. The USRA coauthors gratefully acknowledge NASA funding through contract NNM13AA43C. The UAH coauthors gratefully acknowledge NASA funding from cooperative agreement NNM11AA01A. E.B. and C.M. are supported by an appointment to the NASA Postdoctoral Program, administered by the Universities Space Research Association under contract with NASA. D.K., C.A.W.H., and C.M.H. gratefully acknowledge NASA funding through the \Fermi-GBM project.

\clearpage
\bibliography{bibgrb170817a}

\end{document}